\documentclass{aa}  

\usepackage{graphicx}
\usepackage{txfonts,textcomp}

\usepackage[colorlinks=true, allcolors=blue]{hyperref}
\usepackage{orcidlink}

\begin{document} 

    \title{Active nests at solar maximum: How nested flux emergence dominated flaring activity and structured the heliosphere}

   \titlerunning{Active nests at solar maximum}
   \authorrunning{Finley et al.}

   \author{A. J. Finley~\orcidlink{0000-0002-3020-9409}\inst{1},
          A. S. Brun~\orcidlink{0000-0002-1729-8267}\inst{2},
          A. Strugarek~\orcidlink{0000-0002-9630-6463}\inst{2},
          B. Perri~\orcidlink{0000-0002-2137-2896}\inst{2},
          D. M\"{u}ller~\orcidlink{0000-0001-9027-9954}\inst{1},
          M. Janvier~\orcidlink{0000-0002-6203-5239}\inst{1},\\
          A. S. H. To~\orcidlink{0000-0003-0774-9084}\inst{1},
          H. Eklund~\orcidlink{0000-0001-9597-3726}\inst{1}
          }

        \institute{European Space Agency, ESTEC, Noordwijk, The Netherlands\\ \email{adam.finley@esa.int} 
        \and
        Universit\'e Paris-Saclay, Universit\'e Paris Cit\'e, CEA, CNRS, AIM, 91191, Gif-sur-Yvette, France\\ 
        }

   \date{Received July 30, 2026; accepted -- --, 202-}

\abstract 
{Multi-viewpoint observations from near-Earth satellites and ESA’s Solar Orbiter enable a global view of the Sun. This continuous coverage is essential for characterising active nests, prolific sources of solar activity that survive multiple solar rotations.}
{We investigated how the formation of active nests during the maximum of solar cycle 25 affected the Sun’s flaring activity and large-scale magnetic field.}
{Active nests were identified from their persistent activity in Carrington coordinates. For each nest, we monitored the evolution of its photospheric magnetic field, extreme ultraviolet emission, and associated solar flares. Their impact on the solar corona was explored through potential field source surface extrapolations and coronagraph observations. These results were then compared with in situ measurements of solar wind speed and magnetic field polarity.}
{During 156 days of global monitoring in 2024, nearly 80\% of all solar flares originated from three active nests in the southern hemisphere. Two of the nests converged over several months, colliding with a burst of activity in October 2024. The third nest was linked to the emergence of NOAA active region AR13664 in May 2024. Magnetically complex NOAA active regions (classified as $\beta\gamma$, $\beta\delta$, or $\beta\gamma\delta$) were more prevalent within active nests. In 2024, 28\% of the 147 nested active regions were complex, compared to just 16\% of the 181 other regions. In January 2025, two active nests re-emerged from the collision site, and a new active nest appeared in the north. This configuration reinforced a tilted dipole topology in the solar corona throughout 2025.}
{Active nests modulated global solar activity in 2024 and shaped the solar corona and wind in 2025. Nearly continuous tracking was used to isolate their flaring and flux emergence rates from the rest of the Sun. Active nest migration was driven by the emergence of new magnetic flux, implying a coherent sub-surface origin. By anchoring the coronal magnetic field, active nests created more favourable source regions for solar wind connectivity studies.}


   \maketitle
%

\section{Introduction}

The Sun reached the maximum of its 11-year activity cycle in 2024 and has since transitioned into the declining phase \citep{miesch2025solar, jouve2025forecasting}. The solar cycle is sustained by global dynamo action \citep[for a review, see][]{brun2017magnetism}, which drives magnetic flux from the deep interior to emerge at the surface, forming active regions. As the cycle progresses, the emergence latitude of active regions migrates towards the equator, and the Sun's polar magnetic fields reverse \citep{hathaway2011standard}. Active regions also tend to emerge preferentially at certain Carrington longitudes \citep{berdyugina2003active, mandal2017solar, karapinar2026quantifying}, shaping the distribution of observed solar flares \citep{pojoga2002clustering, temmer2005causes, gyenge2016active, loumou2018association}. When multiple active regions emerge in close proximity, they create long-lived sources of solar activity called `active nests' \citep{harvey1993properties, schrijver2008solar, norton2025moderate}. Active nests persist for several months, far exceeding the lifetimes of individual active regions, which typically endure for only a few weeks \citep[][]{schmieder2014magnetic}. Furthermore, these nests contribute substantially to the total number of solar eruptions \citep[][]{korsos2025solar}. Although the exact physical mechanism responsible for active nest formation is uncertain, they must be connected to how the solar dynamo generates and stores magnetic field in the deep interior \citep{nelson2012magnetic, strugarek2023dynamics}, how buoyant flux tubes rise through the convection zone \citep{jouve2018interactions, dikpati2021deciphering}, and how magnetic flux emerges through the photosphere \citep{chen2017emergence, toriumi2024convective}.

Active nests build complex coronal magnetic fields with enough free magnetic energy to sustain flaring activity over several solar rotations \citep{blaise2026nlff}. At large scales, active nests modify the solar wind outflow by supporting large closed magnetic loops that anchor the heliospheric current sheet (HCS) above \citep{balthasar1983preferred, neugebauer2000solar, benevolenskaya2005formation, yang2024observing, finley2024nested, tahtinen2024straight}. The combination of stored free magnetic energy and large-scale connectivity could make active nests highly susceptible to sympathetic eruptions \citep[e.g.][]{guite2025flaring, wang2026major}. In addition, large patches of unipolar magnetic flux from decaying active nests form coronal holes \citep{mordvinov2014reversals} and interact with pre-existing magnetic flux to create filament channels \citep{gaizauskas2008development, finley2025prolific}. Multiple active nests can coexist on the Sun at any given time, typically appearing in pairs separated by 180$^{\circ}$ in longitude \citep{bumba2000longitudinal, berdyugina2003active, usoskin2005preferred, mandal2017solar}. Consequently, characterising their individual behaviours and mutual interaction \citep[e.g.][]{gaizauskas1983large} is critical for improving short- to medium-term solar activity forecasting. Multi-viewpoint studies incorporating far-side data from ESA’s Solar Orbiter \citep{muller2020solar} are essential to this endeavour \citep[e.g.][]{kontogiannis2025near}, as this enables nearly continuous monitoring of the entire solar surface for several months each year \citep{zouganelis2020solar}. However, comprehensive long-term 360$^{\circ}$ studies of active nests remain scarce.

In this study, we combined observations from near-Earth satellites with far-side data from Solar Orbiter to characterise the flaring activity, morphology, and magnetic field evolution of active nests during the maximum of solar cycle 25. Section~\ref{sec:2} details the data products used in this analysis. In Sect.~\ref{sec:3} we summarise the solar activity in 2024 as seen by near-Earth satellites, identify three prominent active nests, and leverage far-side observation from Solar Orbiter to track these active nests near-continuously from April to October. Section~\ref{sec:4} explores the evolution of each active nest and quantifies their relative contributions to global solar activity in 2024, highlighting patterns in their flux emergence, active region complexity, and flare production. In Sect.~\ref{sec:5} we discuss the influence that active nests had on the Sun's large-scale magnetic field and the structure of the heliosphere in 2025. We conclude that short- to medium-term space weather forecasting could be improved by incorporating these observational trends in activity and large-scale magnetic field organisation.

\section{Data products}\label{sec:2}

\begin{figure}[t]
    \centering
    \includegraphics[trim=0cm 0cm 0cm 0cm, clip, width=0.5\textwidth]{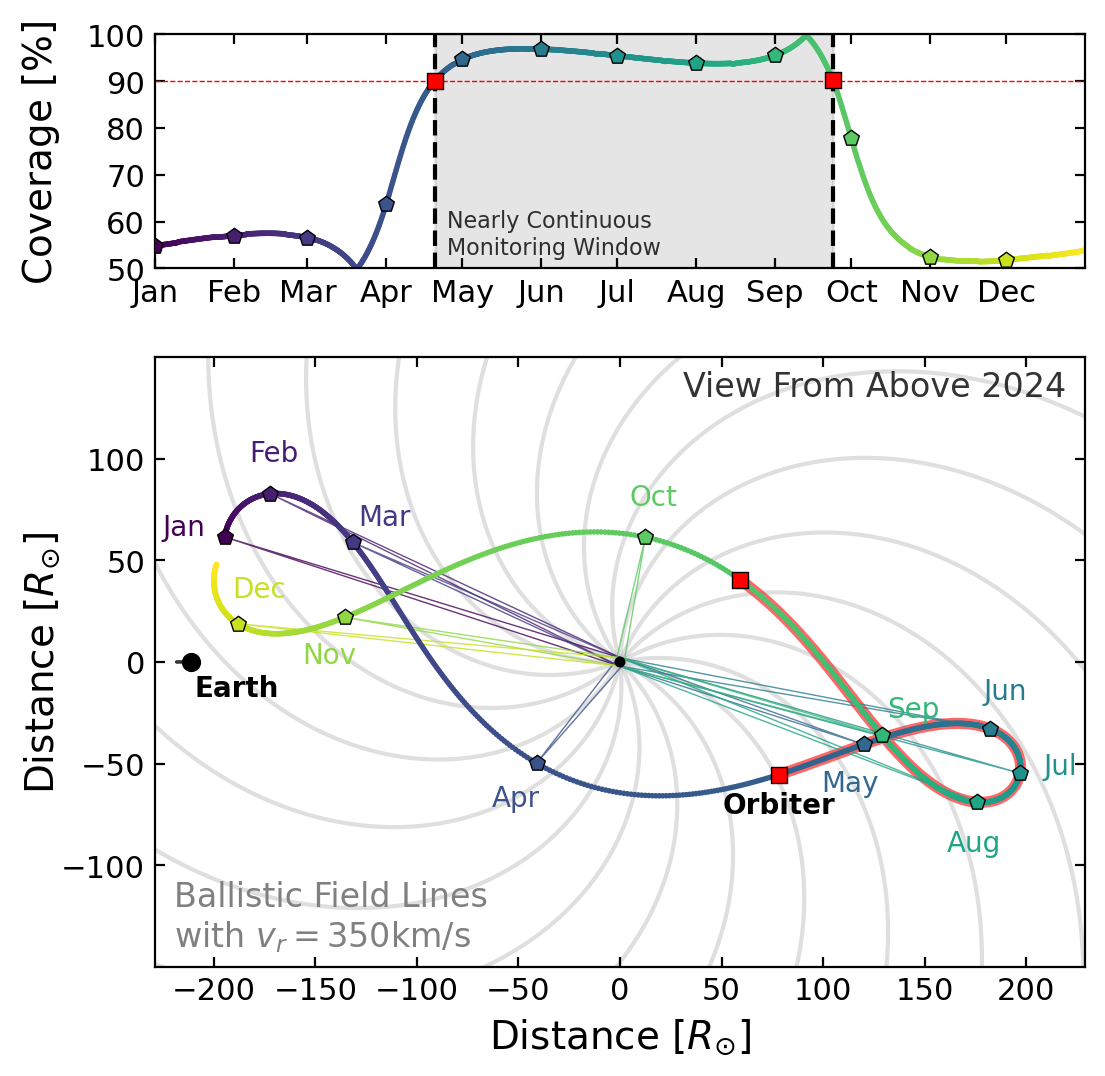}
    \caption{Summary of the multi-viewpoint observations from 2024. \textit{Top}: Solar surface area coverage obtained from the combined viewpoints. \textit{Bottom}: Solar Orbiter's position in the Earth-Sun reference frame. A threshold of 90\% coverage was used to define the nearly continuous monitoring window.}
    \label{fig:soloOrbit}
\end{figure}

Multi-viewpoint observations are essential for monitoring long-lived features, due to the Sun's rotation relative to the Earth. The methodology used in this study follows the approach established by \citet{finley2025prolific} for the nearly continuous monitoring of active nests during the solar cycle's rising phase in 2022. We combined observations from the Solar Dynamics Observatory \citep[SDO;][]{pesnell2012solar} and the Geostationary Operational Environmental Satellite \citep[GOES;][]{woods2024goes} system with their counterparts from Solar Orbiter. The relative positions of Solar Orbiter and the Earth in 2024 are displayed in Fig. \ref{fig:soloOrbit}. Full-disk images at extreme ultraviolet (EUV) wavelengths -- specifically the 304~\AA\  channel -- were taken from the Atmospheric Imaging Assembly \citep[AIA;][]{lemen2012atmospheric} on SDO and the Extreme Ultraviolet Imager \citep[EUI;][]{rochus2020solar} on board Solar Orbiter. Carrington maps of EUV emission were constructed using the SunPy package \citep{sunpy_community2020} to re-project and combine multiple full-disk observations using a weight kernel that favoured observations closer to the sub-solar point. 

Line-of-sight magnetic field observations were taken from the Helioseismic and Magnetic Imager \citep[HMI;][]{scherrer2012helioseismic} on SDO and the Polarimetric and Helioseismic Imager \citep[PHI;][]{solanki2020polarimetric} on board Solar Orbiter. Previous works have compared the magnetic field strengths recovered by HMI and PHI \citep[e.g.][]{loeschl2024first} and inter-calibrated the two instruments \citep{sinjan2023magnetic, vacas2024comparison}. For this study, we evaluated the unsigned magnetic flux in regions of interest as they crossed the central meridian using the full-disk observations from each observer. PHI magnetic field strengths were multiplied by 1.3; this factor was derived by matching the magnetic flux in active regions observed simultaneously by HMI near the central meridian at the beginning and end of 2024. 

Flare statistics from GOES were retrieved from the Heliophysics Event Knowledgebase \citep[HEK;][]{hurlburt2012heliophysics}, while the corresponding information was compiled from the Spectrometer/Telescope for Imaging X-rays \citep[STIX;][]{krucker2020spectrometer} on board Solar Orbiter. To generate the list of STIX flares, we set a threshold of 1000 counts s$^{-1}$cm$^{-2}$ in the $4-10$ keV energy band. Flare locations were then estimated using the xrayvision\footnote{\href{https://github.com/TCDSolar/xrayvision}{https://github.com/TCDSolar/xrayvision}} python package. This methodology is identical to the STIX science flare list\footnote{\href{https://github.com/hayesla/stix\_flarelist\_science}{https://github.com/hayesla/stix\_flarelist\_science}}. To assign GOES flux classifications (e.g. C, M, X, and X10) to the STIX flares, we converted the STIX peak counts $N_{STIX}$, normalised by spacecraft distance $R$ in astronomical units, to an approximate GOES 1.5 - 12.4 keV peak flux $F_{GOES}$ using a power-law scaling,
\begin{equation}
    \log_{10}(F_{GOES}) = a + b\cdot\log_{10}(N_{STIX} \cdot R^2).
    \label{eq:flare}
\end{equation}
Similar to \citet{xiao2023data} and \citet{finley2025prolific}, the parameters of $a=-7.1$ and $b=0.46$ were fit to flares observed by both instruments at the beginning and end of 2024 (see Appendix \ref{ap:goes_stix}). Flares detected by both GOES and STIX with peak times within five minutes of each other and that were located within 5$^{\circ}$ in Carrington coordinates were combined. We used the GOES flux classification from the observer with the best viewing angle. Large M- and X-class flares triggered the STIX attenuator to be inserted, which reduced their recorded counts. We did not correct for the attenuator in our assigned GOES classifications; instead, we analysed these flares separately. In future, these classifications could be corrected using the STIX background detector, as it is not covered by the attenuator \citep[see][]{stiefel2025using}.

\begin{figure*}
    \centering
    \includegraphics[trim=0cm 0cm 0cm 0cm, clip, width=\textwidth]{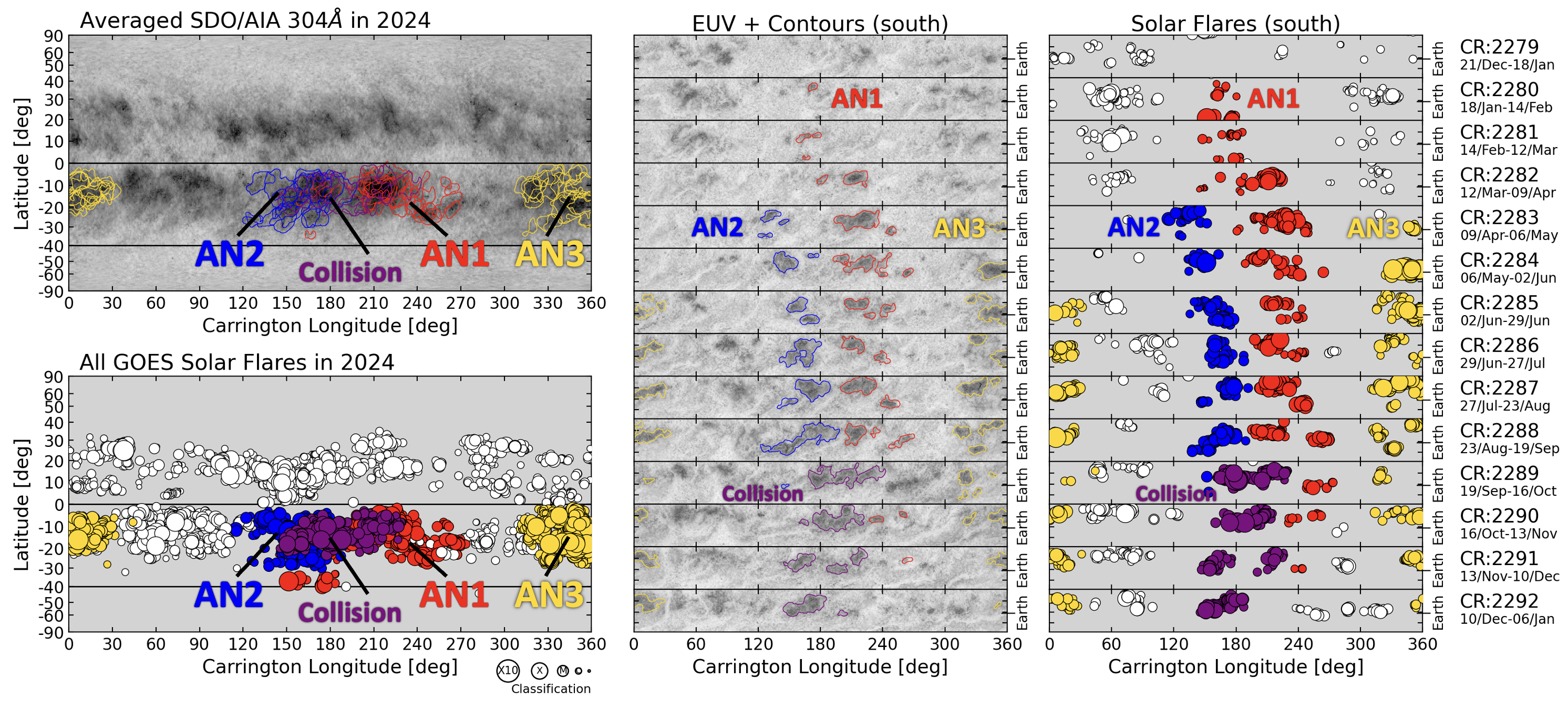}
    \caption{Overview of solar activity in Carrington coordinates for 2024. \textit{Top left}: Averaged SDO/AIA 304~\AA\  emission. \textit{Bottom left}: Locations of all solar flares observed by GOES. \textit{First column}: Time-evolution of EUV emission in the southern hemisphere. \textit{Second column}: Time-evolution of the solar flares. Time progresses from top to bottom in both columns. The EUV emission and flares associated with active nests AN1, AN2, and AN3 are coloured red, blue, and yellow, respectively. The collision product of AN1 and AN2 is coloured purple. }
    \label{fig:CRsummary}
\end{figure*}

\section{Identification and tracking of active nests}\label{sec:3}
\subsection{Active nests viewed from Earth}\label{sect:nests}

We investigated the role of active nests during the 2024 solar activity maximum. The distribution of solar activity in 2024 is displayed in Fig. \ref{fig:CRsummary}, which contains Carrington rotations (CRs) 2279 to 2292. To identify active nests, we used the averaged EUV emission from AIA 304~\AA\ Carrington maps and the locations of GOES flares in Carrington coordinates. There was a distinct north-south asymmetry in solar activity during 2024, with $\sim$70\% of all GOES flares (mostly C- and M-class) originating in the southern hemisphere. Three regions in the southern hemisphere exceeded the 90th percentile of the averaged EUV emission, denoted AN1, AN2 and AN3. The GOES flares were strongly concentrated within these regions. AN1 and AN2 migrated towards each other from 240$^{\circ}$ and 120$^{\circ}$ Carrington longitude, respectively, while AN3 was on the opposite side of the Sun, near to 355$^{\circ}$. We focused on these three active nests; however, this is not a unique selection. Notably, activity nesting was present for a few CRs at the start and end of 2024 in the southern hemisphere around 60$^{\circ}$ longitude and in the northern hemisphere drifting from 150$^{\circ}$ to 180$^{\circ}$ and then 130$^{\circ}$ longitude during 2024.

The morphological evolution of AN1, AN2, and AN3 was extracted using the EUV emission from AIA, before the process was repeated for the far-side data. For each CR in 2024, brightness contours were extracted from AIA 304~\AA\ Carrington maps following the application of a Gaussian smoothing. The absolute size of each contour was slightly affected by the local environment of each feature. The selection threshold was intentionally set to capture the full extent of features, at the cost of including some of the surrounding quiet Sun. Priority was given to continuity in Carrington longitude, following the assumption that the initial rise of flux tubes through the convection zone may be less coherent and thus slightly spread in latitude. For example, in CRs 2280 and 2281, AN1 includes some flux emergence events towards the south of the main nesting site. GOES flares were then assigned to their nearest contour. Finally, we manually assigned the contours and their associated flares to AN1, AN2, and AN3 based on their Carrington longitude to highlight regions of persistent solar activity. By October, the proximity of AN1 and AN2 meant that they were no longer separable. This combined region was labelled as a product of their collision. The longitudinal evolution of the active nests is illustrated in Fig. \ref{fig:CRsummary}. Each active nest sustained coherent activity that spanned several solar rotations, satisfying the definition from \citet{finley2025prolific}.

\subsection{Nearly continuous monitoring with Solar Orbiter}\label{sect:orbiter}

Solar Orbiter's trajectory provides opportunities to observe the Sun's far-side with respect to Earth. When combined with data from near-Earth satellites (e.g. SDO and GOES), this creates epochs with nearly continuous monitoring of the entire solar surface. The relative position of Solar Orbiter to Earth during 2024 is displayed in Fig. \ref{fig:soloOrbit}. From April to October, Solar Orbiter had a unique view of the Sun's far-side that complemented near-Earth satellites. This allowed us to perform nearly continuous monitoring of the three active nests (AN1, AN2, and AN3). When assessing global statistics, we required at least 90\% of the solar surface to be visible. This threshold produced a nearly continuous monitoring window of 156 days, from 20 April to 24 September. Venus gravity assist manoeuvres in February 2025, December 2026, and so on, will increase the spacecraft's orbital inclination, and reduce the duration of these nearly continuous monitoring windows from $\sim$150 days per year to an average of $\sim60$ days (alternating each year after 2026 between 25 and 90 days). 

The same feature-extraction process used on AIA was applied to the EUI 304~\AA\ Carrington maps. To ensure consistency with AIA, the threshold was manually adjusted to reproduce the contours from AIA at times when the two instruments observed the same features. Full-disk EUV images from AIA and EUI during the nearly continuous monitoring window are shown in Appendix \ref{ap:nestCollisionView}, along with the derived contours for AN1 and AN2. In October, the two active nests began to interact as Solar Orbiter left the Sun's far-side and the nearly continuous monitoring window ended. To complete the flare distribution, events detected by STIX were assigned to AN1, AN2, or AN3 based on their reconstructed Carrington coordinates, as done previously for flares detected by GOES. The result is presented in Fig. \ref{fig:EarthPlusOrbiter}. The nearly continuous monitoring window captured the complete evolution of AN2 and its collision with AN1; however, we lacked some coverage at the beginning and end of the window for AN1, AN3, and the collision product.

\section{Active nest evolution}\label{sec:4}
\subsection{Contribution to global flaring activity}

After tracking the active nests, we evaluated their contribution to global solar activity. Figure \ref{fig:global_flares} summarises the solar flare occurrence rate as a function of flare classification during 2024. Following the increased coverage of the solar surface (see Fig. \ref{fig:soloOrbit}), the number of flares observed within the nearly continuous monitoring window grew by 35–40\% compared to when Solar Orbiter was closer to Earth. The overall level of solar activity was consistent with 2024 being the maximum of solar cycle 25, with a total of 11,533 C-class, 1,260 M-class, and 37 X-class flares observed by GOES and STIX (plus 73 events that triggered the STIX attenuator). While periods of heightened M- and X-class activity appeared every few months, the southern hemisphere consistently hosted more flaring activity. Over the full year, we attributed 74\% of C-class, 78\% of M-class, and 78\% of X-class flares to the southern hemisphere. 

The north-south asymmetry in flaring activity correlates with the lag in sunspot number between hemispheres \citep{joshi2015evolutionary}. Since solar cycle 20, the southern hemisphere has been trailing behind the northern hemisphere \citep{temmer2002hemispheric}. Cycles 23 and 24 had a two-year delay between the peak sunspot number in each hemisphere, following the emergence of a strong quadrupole mode \citep{derosa2012solar, finley2023evolution}. A similar trend has also emerged for cycle 25. The quadrupole-to-dipole energy ratio reached a maximum of $\sim40$ near the end of 2023, and remained around $\sim5$ throughout 2024, before the large-scale field became strongly influenced by the dipole component (discussed in Sect. \ref{sec:5}). Accordingly, during the 156 days of continuous global coverage with Solar Orbiter in 2024, flaring activity was dominated by the southern hemisphere, which produced 76\%, 78\%, and 77\% of all C-, M-, and X-class flares, respectively. During the maximum and declining phases of cycles 23 and 24, flare production was similarly concentrated in the southern hemisphere \citep{abdel2018study, korsos2025solar}. 

\begin{figure}
    \centering
    \includegraphics[trim=0cm 0cm 0cm 0cm, clip, width=0.5\textwidth]{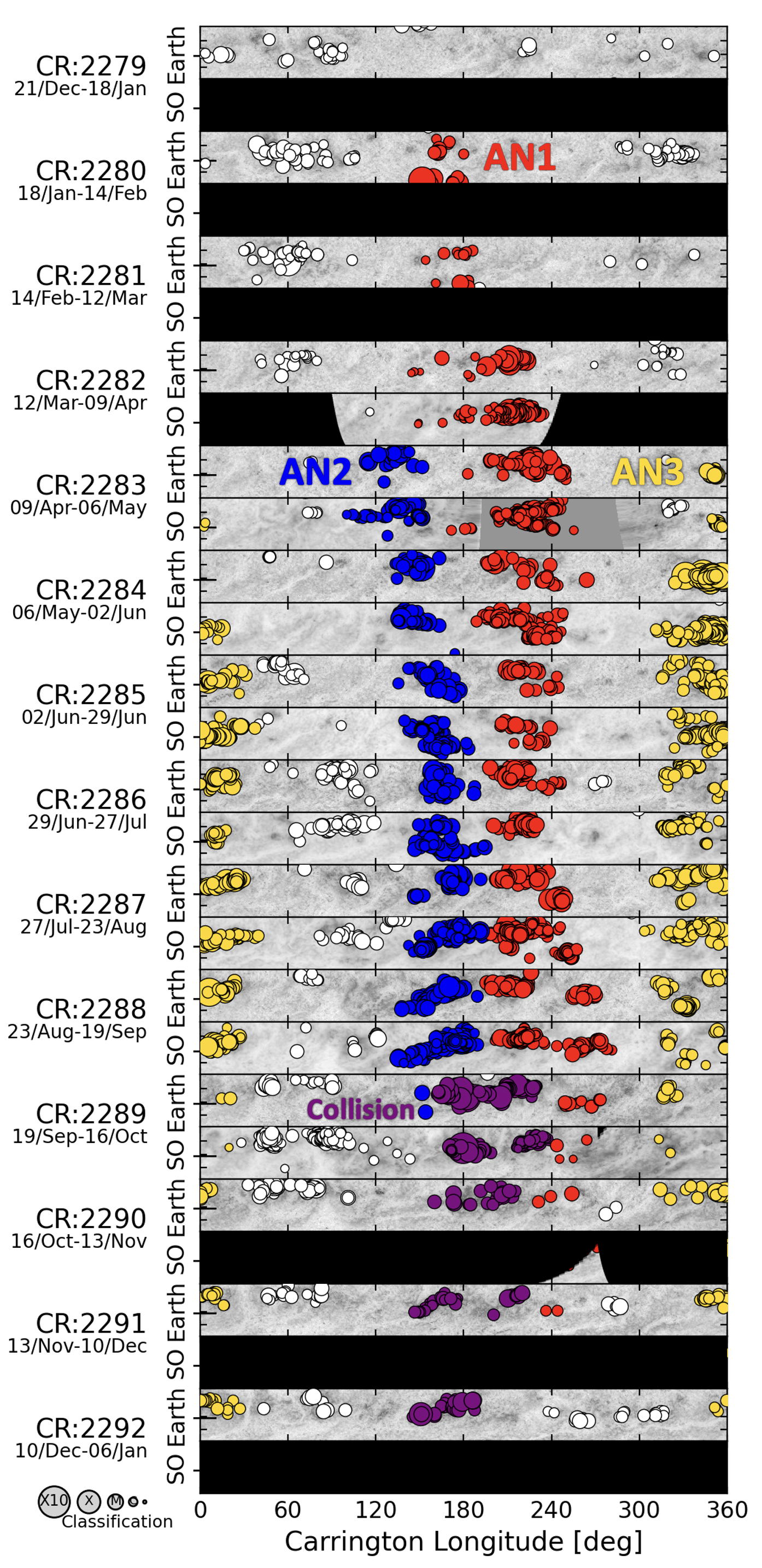}
    \caption{Longitudinal distribution of EUV emission (background grey scale) and solar flares (circle markers) in the southern hemisphere in 2024. Each CR is split to show data from near-Earth and Solar Orbiter. When the observations from Solar Orbiter provide no additional information, the data are masked by black bars.}
    \label{fig:EarthPlusOrbiter}
\end{figure}

\begin{figure*}[h!]
    \centering
    \includegraphics[trim=0cm 0cm 0cm 0cm, clip, width=0.9\textwidth]{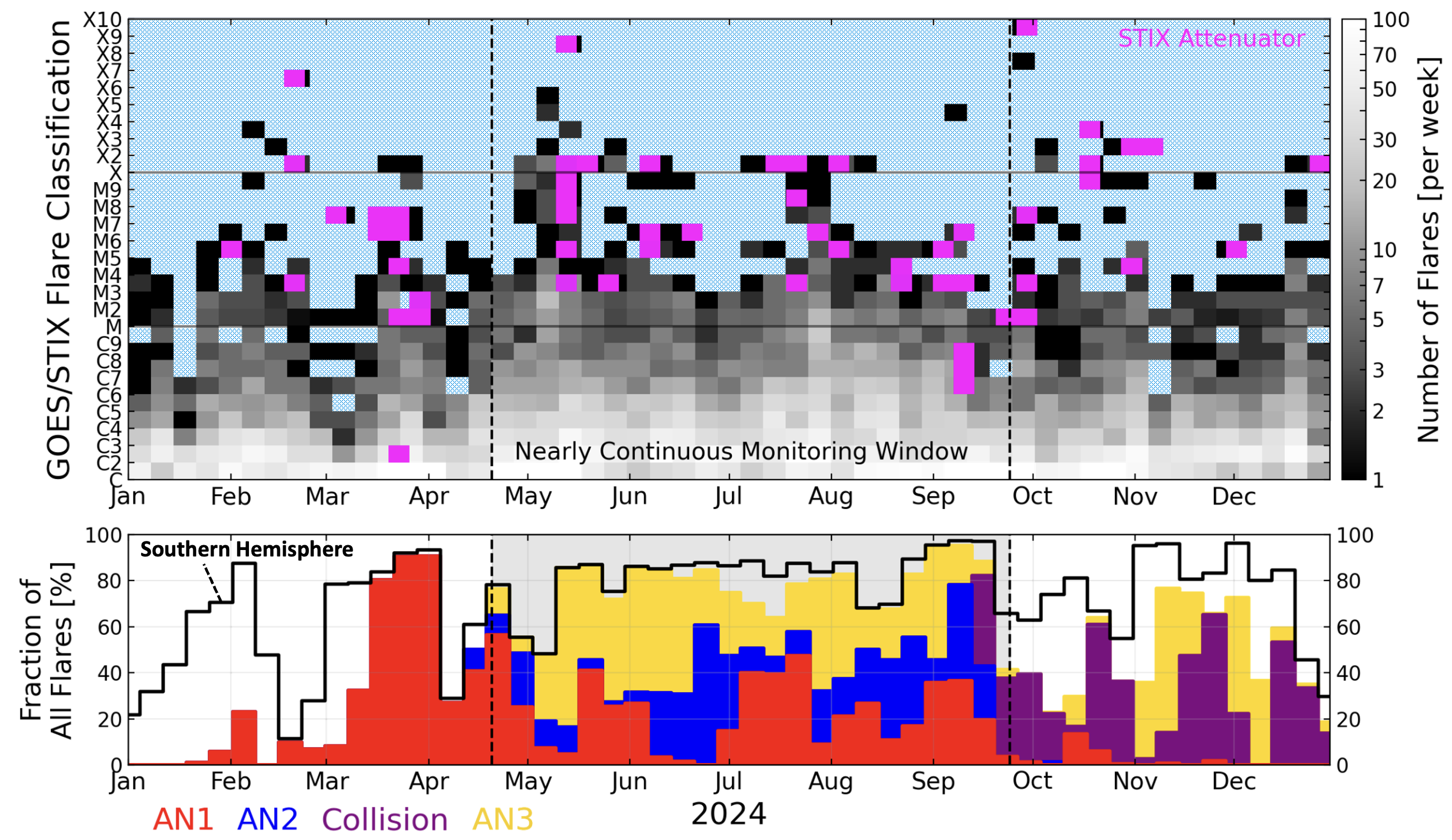}
    \caption{Summary of solar flares observed by GOES and STIX in 2024. \textit{Top}: Histogram of all flares as a function of classification and time. Flares that triggered the insertion of the STIX attenuator are highlighted in magenta; their classifications have not been corrected. \textit{Bottom}: Fraction of flares occurring within AN1, AN2, and AN3, irrespective of classification. The total fraction of flares originating in the southern hemisphere is indicated by the thick black line. Vertical dashed lines mark the bounds of the nearly continuous monitoring window between Earth and Solar Orbiter (with at least 90\% of the solar surface visible). }
    \label{fig:global_flares}
\end{figure*}

\begin{figure}[h!]
    \centering
    \includegraphics[trim=0cm 0cm 0cm 0cm, clip, width=0.5\textwidth]{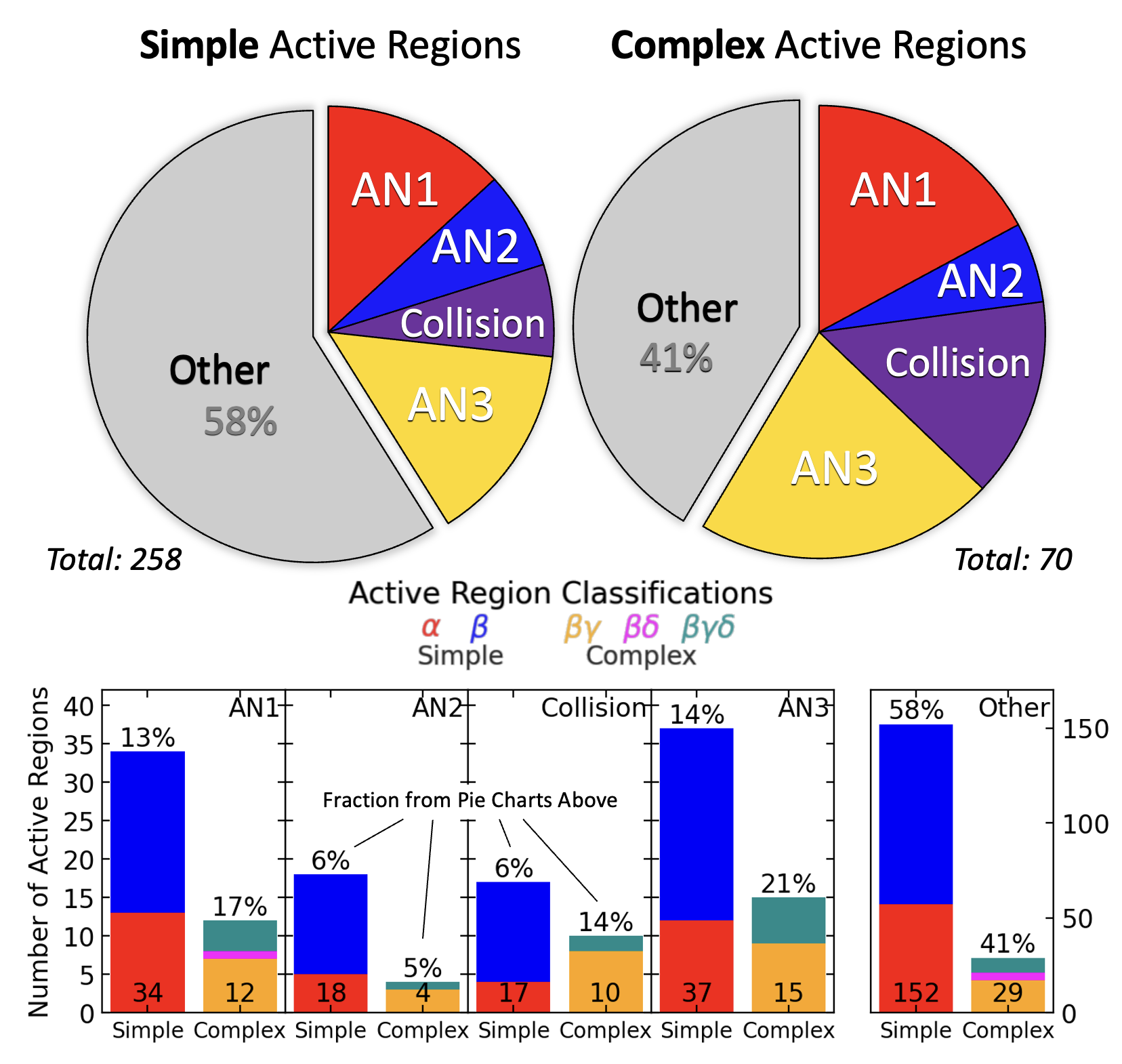}
    \caption{Fraction of flaring NOAA active regions in 2024 with simple ($\alpha$ and $\beta$) or complex ($\beta\gamma$, $\beta\delta$, and $\beta\gamma\delta$) magnetic classifications. \textit{Top}: Total number of simple and complex active regions located within nests versus elsewhere. \textit{Bottom}: Detailed breakdown of $\alpha$, $\beta$, $\beta\gamma$, $\beta\delta$, and $\beta\gamma\delta$ active regions inside each active nest compared to those outside of nests.}
    \label{fig:complexity}
\end{figure}

\begin{figure*}[h!]
    \centering
    \includegraphics[trim=0cm 0cm 0cm 0cm, clip, width=\textwidth]{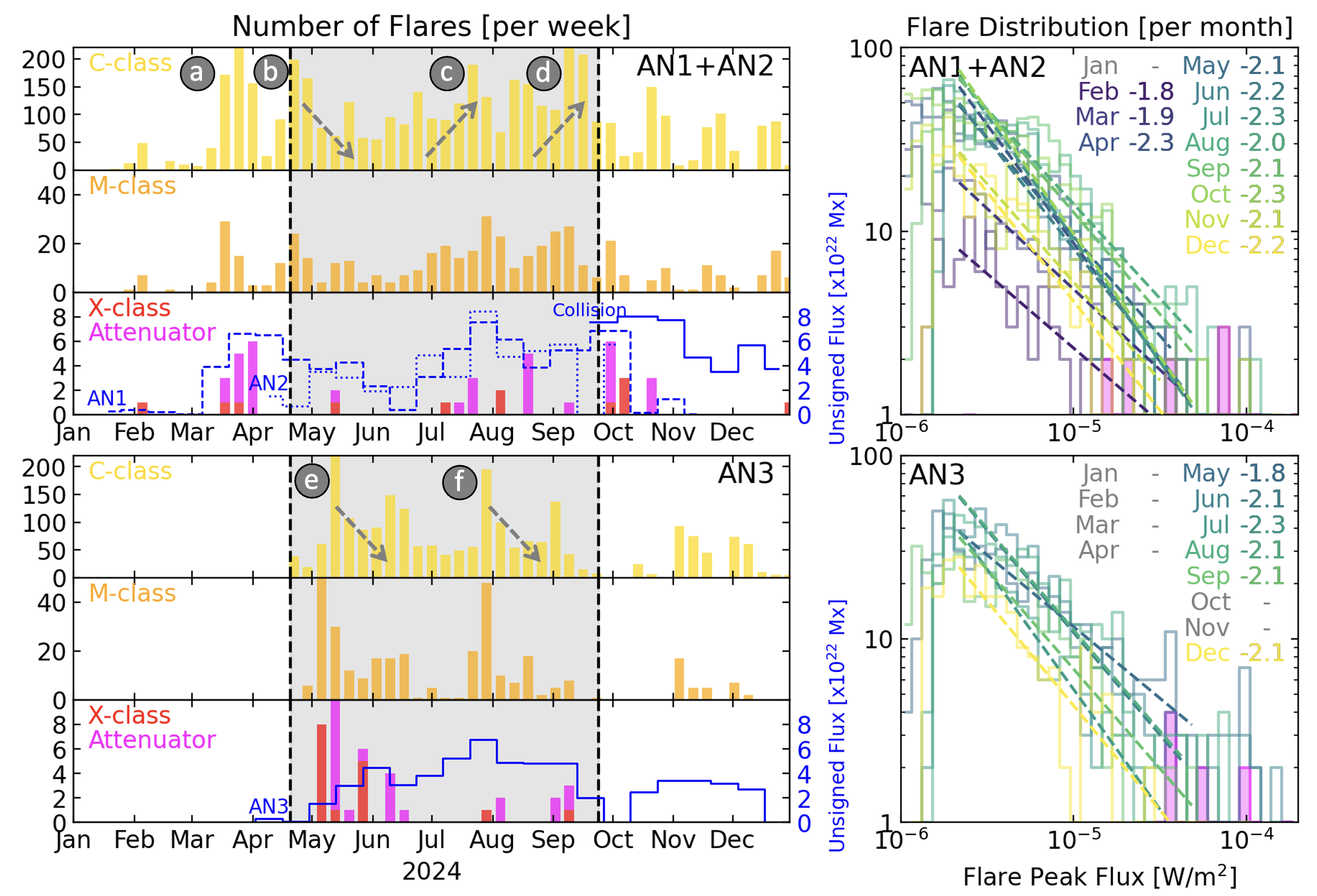}
    \caption{Evolution of flaring activity observed by GOES and STIX for the active nests in 2024, with values for AN1 and AN2 combined. \textit{Left}: Number of flares in each GOES class. Flares affected by the STIX attenuator are plotted alongside X-class events, though the former likely include strong M-class events. Epochs with elevated activity are denoted (a) to (f). Dashed arrows highlight steadily increasing or decreasing activity. The unsigned magnetic flux for each active nest (blue lines) was calculated during central meridian crossings for both SDO and Solar Orbiter. The light grey background and vertical dashed lines denote the window of nearly continuous monitoring. \textit{Right}: Power-law fits to the monthly flare distributions as a function of peak flux.}
    \label{fig:magnetic_flux}
\end{figure*}

\begin{figure*}[h!]
    \centering
    \includegraphics[trim=0cm 0cm 0cm 0cm, clip, width=\textwidth]{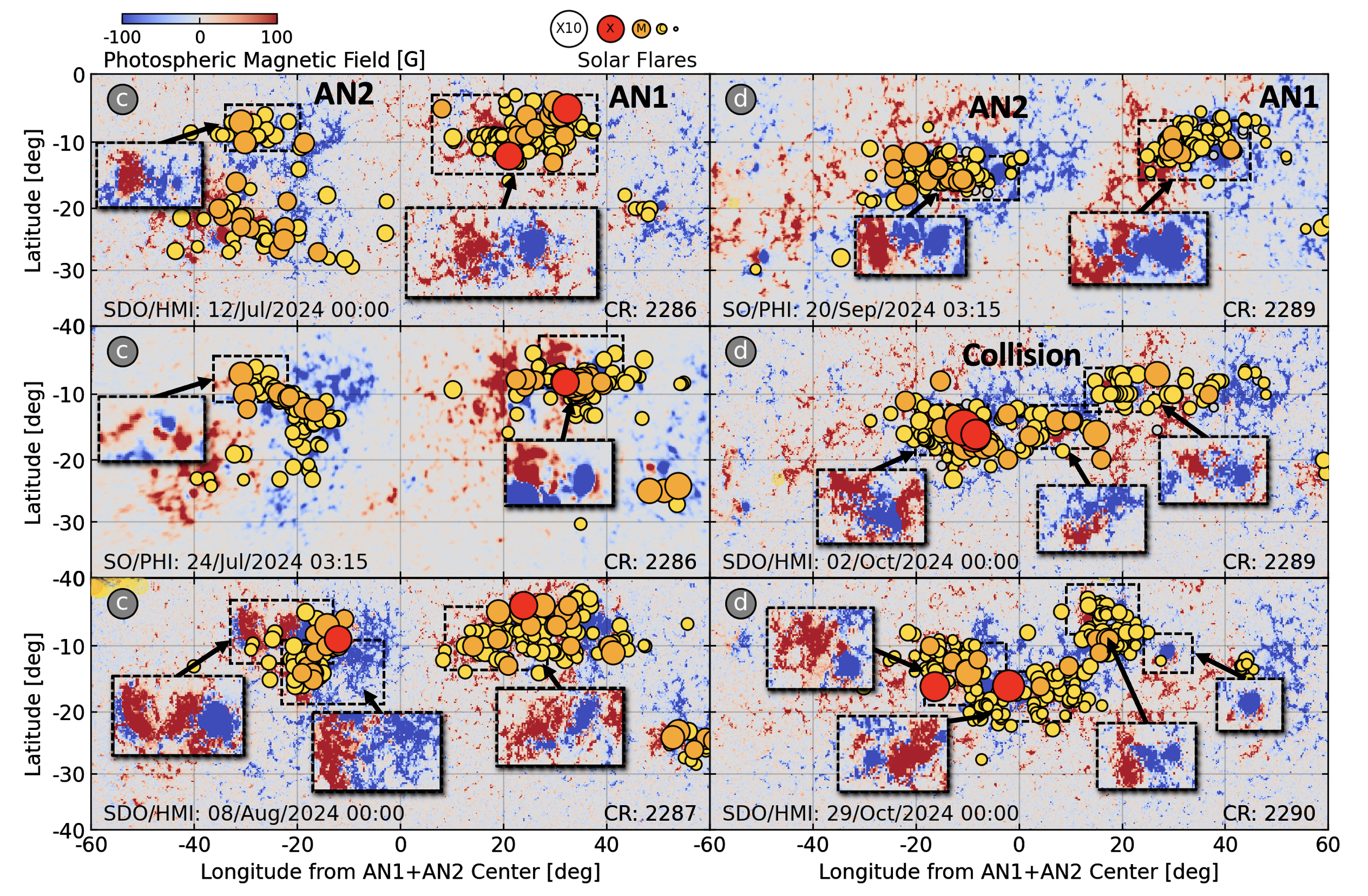}
    \caption{Magnetic flux emergence events within AN1 and AN2 from July to August (\textit{left}) and September to October (\textit{right}) that correspond to the increased activity periods denoted by (c) and (d) in Fig. \ref{fig:magnetic_flux}. Each photospheric magnetogram was captured when the midpoint between the two nests (190$^{\circ}$ Carrington longitude) crossed the central meridian from the perspective of either Earth or Solar Orbiter. The locations of flares occurring within one week of each magnetogram are overplotted, with marker sizes and colours scaled by their GOES classifications. Distinct flux emergence events in each period are highlighted within the inset boxes.}
    \label{fig:collision}
\end{figure*}

\begin{figure}[h!]
    \centering
    \includegraphics[trim=0cm 0cm 0cm 0cm, clip, width=0.45\textwidth]{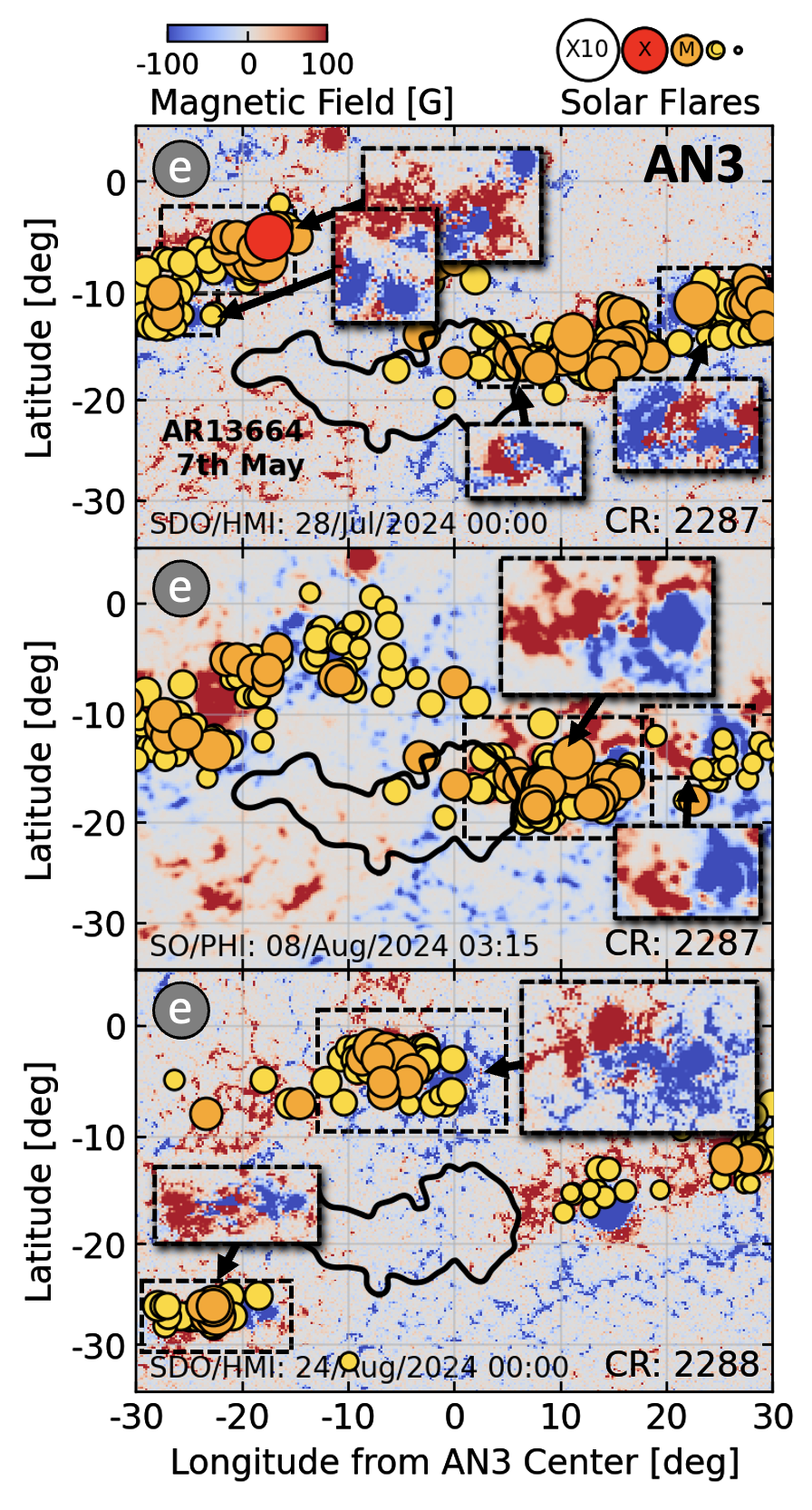}
    \caption{Same as Fig. \ref{fig:collision} but for the burst of activity in AN3 denoted by (e) in Fig. \ref{fig:magnetic_flux}. The black contour outlines the emerged magnetic field from NOAA AR13664 on 7 May 2024.}
    \label{fig:mayregion}
\end{figure}

The majority of flaring activity in 2024 originated in the three active nests. Flares from each nest spanned 30-40$^{\circ}$ in Carrington longitude. AN1 was the first to exhibit significant flaring at the end of March (CR 2280). AN2 and AN3 became visible to Earth in April (CR 2283), with AN3 containing NOAA AR13664 that was responsible for the largest geomagnetic storm so far this cycle \citep[summarised in][]{hayakawa2025solar}. Accordingly, there was a sharp increase in the number of flares observed in May \citep{kontogiannis2025near, zhang2026waiting}, with AN3 producing $60\%$ of all flares globally. The modelling of AR13664 by \citet{jarolim2024magnetic} provides a detailed look at the evolution of the coronal magnetic field during this time. Solar Orbiter observed the most powerful flare of solar cycle 25 so far, originating from AN3 while on the Sun’s far-side, with its strength estimated to correspond to an X12 flare \citep{jaswal2025deconstructing}. 

Subsequent periods of elevated activity occurred at the ends of July and September, both of which involved the interaction of AN1 and AN2. This interaction triggered NOAA AR13842 to release a series of Earth-directed eruptions that interacted with the overlying HCS \citep{temmer2026coronal}. This included the largest flare observed from Earth so far this cycle \citep[X9.0;][]{ding2025giant}. Solar Orbiter returned towards Earth in October, at which point the Sun's rotation modulated activity between the AN1--AN2 collision product and AN3 separated by $\sim$180$^{\circ}$ longitude (see Fig. \ref{fig:EarthPlusOrbiter}). Solar activity remained concentrated within these two southern nesting sites throughout November. As rotation brought each nest into view of Earth, they accounted for roughly 70\% of all flares on the solar disk (see Fig. \ref{fig:global_flares}).

\subsection{Magnetic complexity of active regions}

The emergence of magnetic flux within active nests is essential for maintaining coherent activity over several solar rotations. Therefore, active nests are composed of a varying population of individual active regions \citep{bumba1969solar, bumba2000longitudinal}, each with different magnetic topologies. The magnetic complexity of NOAA active regions can be described using the modified Hale classification system \citep{hale1919magnetic, kunzel1965klassifikation}, where $\alpha$ and $\beta$ represent relatively simple configurations, while $\gamma$ and $\delta$ denote increasing magnetic complexity and the presence of shared penumbrae between opposite-polarity umbrae. It is well established that $\beta\gamma\delta$ regions are the primary drivers of the most energetic solar flares \citep{sammis2000dependence}. Furthermore, the temporal lag between the emergence of simple bipolar regions and the appearance of complex active regions during the solar cycle suggests that magnetic complexity is a product of cumulative evolution \citep{jaeggli2016magnetic}. Complex active regions also tend to be larger in size. The 10- to 15-month lag of sunspot area behind the number of large flares \citep{temmer2003does} may further reflect this progressive structuring of the magnetic field. This trend appears most pronounced during odd-numbered cycles, pointing to a potential link with the Sun's 22-year magnetic cycle. 

In Fig. \ref{fig:complexity} we summarise the magnetic classifications of flaring NOAA active regions in 2024. NOAA active regions classifications are updated as they evolve and rotate across the visible solar disk, so we selected the classification associated with the largest GOES flare produced by each active region. The resulting 258 simple ($\alpha$ and $\beta$) and 70 complex ($\beta\gamma$, $\beta\delta$, and $\beta\gamma\delta$) active regions were then assigned to AN1, AN2, their collision product, AN3, or elsewhere; following the methodology previously used for solar flares. The active nests accounted for 44\% of all 328 NOAA active regions, yet contained 59\% of the 70 complex regions. Active regions were significantly more complex within the active nests (27.9\% of 147 active regions in nests) than elsewhere on the Sun (16\% of 181 active regions outside of nests). Among the individual active nests, AN2 was closest to the background population with 18\% of 22 active regions being complex. AN1 and AN3 had similar fractions of complex active regions, 26\% of 46 and 29\% of 52, respectively. Strikingly, 37\% of the 27 active regions in the collision region between AN1 and AN2 has complex classifications. While increased complexity within active nests correlated with their prolific flare production, these regions produced a disproportionate number of flares compared to isolated active regions, indicating that nests are more prone to eruption.

\subsection{Active nest flare production}
To track how flaring activity evolved in each active nest, we extracted the frequency of C-, M-, and X-class flares from AN1 plus AN2, and AN3. Figure \ref{fig:magnetic_flux} shows the weekly number of flares along with the unsigned photospheric magnetic flux for each nest. The unsigned flux was evaluated from the photospheric magnetic field within the EUV contour(s) of each nest as they crossed the central meridian for HMI and PHI. We fit the distribution of flares versus peak x-ray flux for each month using a power-law exponent $\alpha$ corresponding to the relation $dN \propto F_X^{\alpha}dF_X$, where $N$ is the flare frequency, and $F_X$ is the flare peak flux (see \citealt{aschwanden2012automated} and references therein). Attenuated STIX flares were excluded from the fitting process, and months with insufficient flare statistics were omitted. The resulting power-law exponent ($\alpha$) ranged from $-1.8$ to $-2.3$. These values are consistent with those from the 2022 active nest studied in \citet{finley2025prolific} as well as previously derived exponents for soft x-ray flares (e.g. \citealt{aschwanden2016review}). 

Overall, AN1, AN2, and their collision product produced 4,466 C-class and 501 M-class flares. AN1 had more activity, responsible for 51\% of these C-class and 49\% of the M-class flares, whereas AN2 accounted for 28\% and 31\%, respectively. The remaining flares were associated with the collision product. M-class flares accounted for 9-10\% of all flares from these regions. In comparison, AN3 produced 2,292 C-class and 325 M-class flares, with 12\% of flares from AN3 reaching M-class. There were a higher number of X-class flares identified in AN3 than in AN1 and AN2 combined (16 versus 12); however, a definitive comparison requires correcting for the STIX attenuator. The elevated GOES background flux during 2024 is expected to have filtered the distribution of smaller C-class flares. The peak unsigned magnetic flux of the individual active nests reached a similar maximum value of around $8\times10^{22}$Mx. 

The flaring activity in each active nest contained distinct epochs of intense activity; annotated in Fig. \ref{fig:magnetic_flux}. These epochs displayed different behaviours, either a steady growth in flaring activity (c and d) or a rapid burst and slow decline (b, e, and f). AN1 was initially active in March (a) and late April (b), with three X-class flares observed by GOES. AN1 and AN2 then produced a steady background of C- and M-class flares in June, building towards late July (c) and September (d). In contrast, both bursts of activity from AN3 in the nearly continuous monitoring window (e and f) appeared to decay over the course of a month. The bursts from AN3 were more pronounced than the peaks in activity from either AN1 or AN2, with the first (e) containing at least nine X-class flares (from NOAA AR13664). The power-law exponent ($\alpha$) was between $-1.8$ and $-2.1$ for each burst in AN3, whereas for AN1 and AN2 $\alpha$ was steeper between $-2.2$ and $-2.3$. In the following subsections, we attribute this difference to how flux emergence was distributed within the active nests.

\subsection{Collision of two nests in October 2024}

The convergence and divergence of active nests have been well documented in previous solar cycles \citep[e.g.][]{pojoga2002clustering}, with interacting nests linked to heightened flaring activity. Notably, the extreme solar activity during the October 2003 `Halloween storms' \citep[see][]{gopalswamy2005coronal} was associated with converging activity in the southern hemisphere. To examine the evolution and collision of AN1 and AN2, Fig. \ref{fig:collision} focuses on the epochs with growing activity denoted (c) and (d) in Fig. \ref{fig:magnetic_flux}. During epoch (c), AN1 and AN2 contained substantial and repeated flux emergence events that were highly spatially correlated with similar bipole orientations. In this way, new magnetic flux emerged directly into pre-existing flux \citep[40-50\% of all active regions emerge like this; e.g.][]{harvey1993properties, norton2025moderate}. This is consistent with the increased magnetic complexity from Fig. \ref{fig:complexity} being produced by the interaction of emerging bipoles. These complex active regions were the primary driver of flaring activity, as evidenced by the clustering of flares within these regions. This may account for the steeper power-law exponent ($\alpha$) since the emergence of new magnetic flux could have destabilised residual structures in the coronal magnetic field, triggering a greater number of small flares relative to large flares, which are more commonly associated with high flux emergence rates \citep{kutsenko2021possibility}. 

Flux emergence events in AN1 and AN2 typically occurred every 12 to 18 days, shifting in longitude by 4$^{\circ}$ to 10$^{\circ}$. As shown in Fig. \ref{fig:collision} for epoch (d), these events were initially confined within individual nests before migrating towards the central region between them. During the collision, the largest eruptions were mostly concentrated within AN2 \citep[e.g.][]{ding2025giant,temmer2026coronal}, following flux emergence events. Figure \ref{fig:magnetic_flux} highlights that the unsigned magnetic flux of the collision product was similar to that of the individual nests (also true of the net magnetic flux). The consistency across nesting sites could be explained by flux emergence events forming from a sub-surface reservoir with a characteristic magnetic field strength. Furthermore, the longitudinal migration of flux emergence events points to a spatially extended source that is progressively depleted over time. Following this hypothesis, AN1, AN2, and the collision product each represent the episodic release of magnetic flux from different sections of the same sub-surface reservoir. The observed collision is therefore not a merger of separate nests, but rather the spatial convergence of flux emergence from a coherent and extended sub-surface structure. This aligns with the global toroidal bands proposed by \citet{dikpati2025mother}, linking surface nesting and sub-surface dynamo reservoirs (as simulated in 3D in \citealp{nelson2014buoyant} and \citealp{jouve2018interactions}).

\subsection{Nesting following the emergence of NOAA AR13664 in May 2024}

Compared to the colliding active nests, AN3 had a distinctly different flux emergence behaviour, with two major episodes of activity, (e) and (f), in Fig. \ref{fig:magnetic_flux}. Epoch (e) corresponds to the rapid emergence of NOAA AR13664, which had a peak emergence rate of $\sim2\times 10^{21}$~Mx/hr \citep[see][]{wang2024unveiling}. Interestingly, the rapid emergence of AR13664 appeared to locally suppress further emergence events, summarised in Fig. \ref{fig:mayregion}. All flux emergence events within AN3 in July and August occurred outside the boundary of emerged flux from AR13664. Additionally, the emergence events to the west and east of AR13664 formed a distinct longitudinal chain. This highly organised pattern supports an extended sub-surface origin for AN3, similar to AN1 and AN2. This is consistent with the global dynamo modelling of \citet{dikpati2025mother} that suggested AR13664 was part of a large-scale toroidal magnetic flux band. With this interpretation, the sub-surface magnetic flux was either exhausted or diverted by the rapid emergence of AR13664. While still complex, the emerging active regions within AN3 were less crowded, which could explain the difference in the power-law exponent ($\alpha$) for the flaring in AN3 with respect to AN1 and AN2. In this case, the separation between active regions allowed them to develop the more characteristic power-law slope, closer to that of isolated active region emergences ($\alpha\approx-1.7$ from \citealp{aschwanden2016review}).

\begin{figure*}[h!]
    \centering
    \includegraphics[trim=0cm 0cm 0cm 0cm, clip, width=\textwidth]{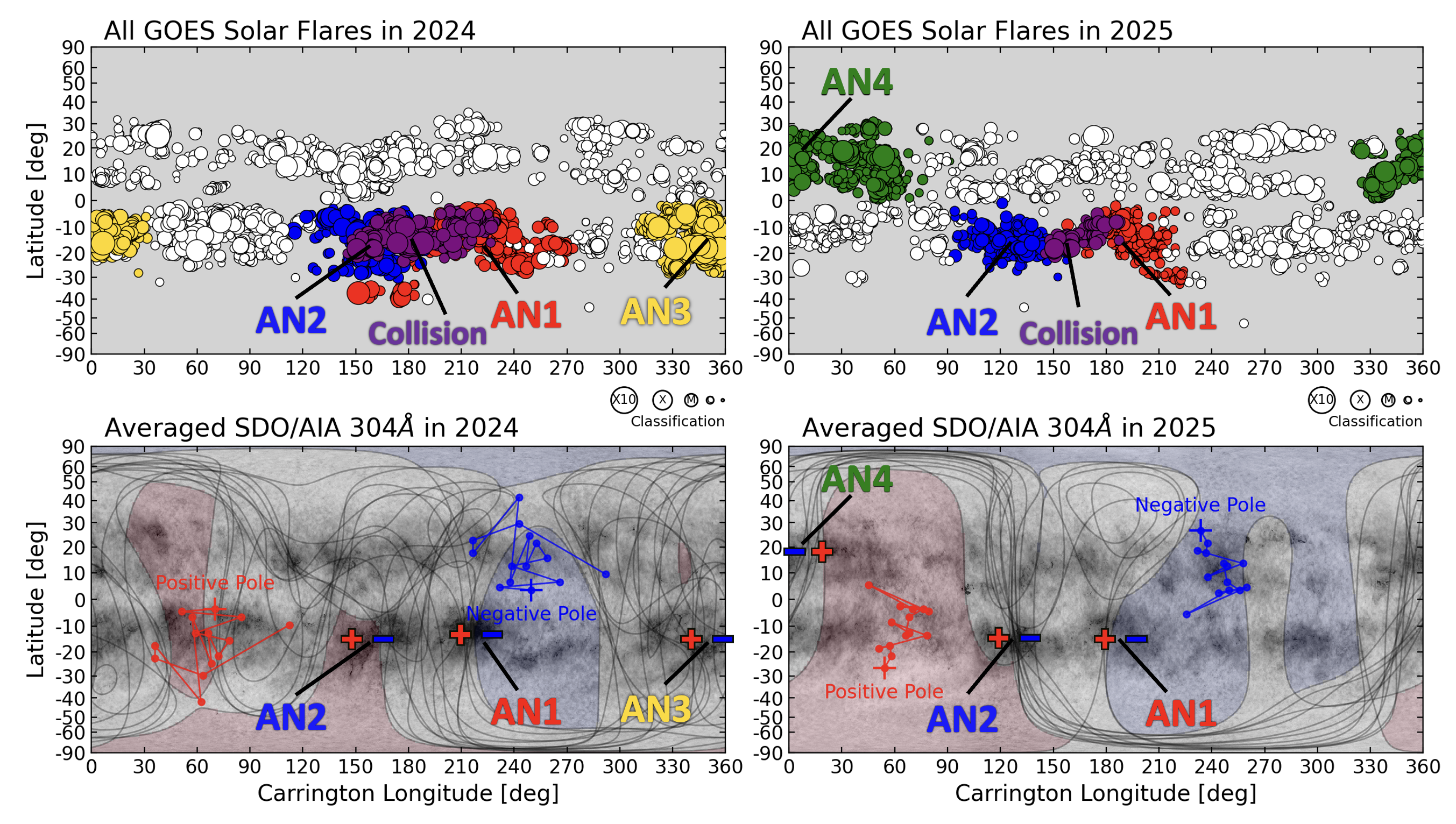}
    \caption{Solar activity and coronal magnetic field polarity in 2024 versus 2025. \textit{Top}: Distribution of all GOES flares from each year, with those assigned to active nests highlighted in colour. \textit{Bottom}: Averaged AIA 304~\AA\  emission in grey scale, overlaid with the coronal magnetic field polarity from SDO/HMI PFSS models for each CR using a source surface height of 2.4 solar radii. Solid black lines represent the HCS from each CR. Light red and blue areas highlight regions of the coronal source surface that maintain a constant magnetic field polarity across all CRs. The evolution of the dipole axis is shown for each year. Plus and minus (red and blue) markers indicate the expected east-west magnetic field orientation for each active nest.}
    \label{fig:2024vs2025}
\end{figure*}

\subsection{Summary of active nests in 2025}

We continued to identify and track nested magnetic activity into 2025, applying the techniques outlined in Sect.~\ref{sec:3}. Two active nests emerged from the remnants of the AN1 and AN2 collision. We retained the labelling conventions of AN1 and AN2 because their spatial configuration mirrored the original nests. The 2025 nests maintained a separation of $\sim120^{\circ}$ in Carrington longitude, though they were offset by $\sim30^{\circ}$ relative to their 2024 positions. Notably, these regions were weaker, reflecting a general decline in solar activity. At the same time, a new active nest (AN4) emerged in the northern hemisphere, near the Carrington longitude of AN3’s final activity. Figure \ref{fig:2024vs2025} illustrates how the distribution of active nests evolved between 2024 and 2025. Areas of high-activity in 2025 somewhat mirrored the quiet regions of 2024, and vice versa. For example, the intense activity observed at 60° longitude in the southern hemisphere in 2024 was completely absent in 2025. Conversely, both hemispheres around 270° longitude exhibited little to no activity in 2024, but these regions then became active in 2025. As this work focused on the intense flaring activity during solar maximum in 2024, a detailed analysis of the 2025 activity was left for future work; however, a brief summary of the 2025 nests (AN1, AN2, and AN4) is provided in Appendix \ref{ap:2025_flares}.

\section{Influence on the heliosphere}\label{sec:5}
\subsection{Coronal magnetic field and solar wind}

The solar wind fills our circumstellar environment, the heliosphere, through which energetic particles and coronal mass ejections propagate \citep[e.g.][]{owens2013heliospheric}. The solar wind is organised by the Sun's large-scale magnetic field \citep{stansby2021active, finley2023evolution}. At solar minimum, the photospheric and coronal magnetic field takes the form of an axisymmetric dipole \citep{mccomas2000solar}. Once active regions emerge, higher-order (quadrupole and octupole) components increasingly affect the Sun's large-scale magnetic field \citep{derosa2012solar}. This produces a more complex solar wind outflow and heliospheric magnetic field at maximum \citep[e.g.][]{reville2017global}. During the solar maximum in 2024, the concentration of flux emergence into active nests had a profound impact on the global structure of the solar corona and wind. 

Nested flux emergence produces coherent structures in the Sun's large-scale magnetic field by reinforcing sectors of opposite polarity that close at higher altitudes in the corona \citep{finley2024nested, tahtinen2026active}. This tends to constrain the HCS, the divide between positive and negative polarity heliospheric magnetic fields, to lie above active nests. We highlight this trend in Fig. \ref{fig:2024vs2025} by computing potential field source surface (PFSS) models \citep{altschuler1969magnetic, wang1992potential, schrijver2003photospheric} for 2024 and 2025 using polar-field-corrected photospheric magnetic field Carrington maps from SDO/HMI \citep{sun2018polar}. In this study, the source surface radius was fixed at 2.4 solar radii for all PFSS models. We identified persistent positive and negative heliospheric magnetic field sectors across all CRs in each year. In 2024, the dipole component of the Sun's large-scale magnetic field was initially weak and inclined to the rotation axis. The dipole strengthened through late 2024 and drifted towards the rotational poles in 2025.

\begin{figure*}[h!]
    \centering
    \includegraphics[trim=0cm 0cm 0cm 0cm, clip, width=\textwidth]{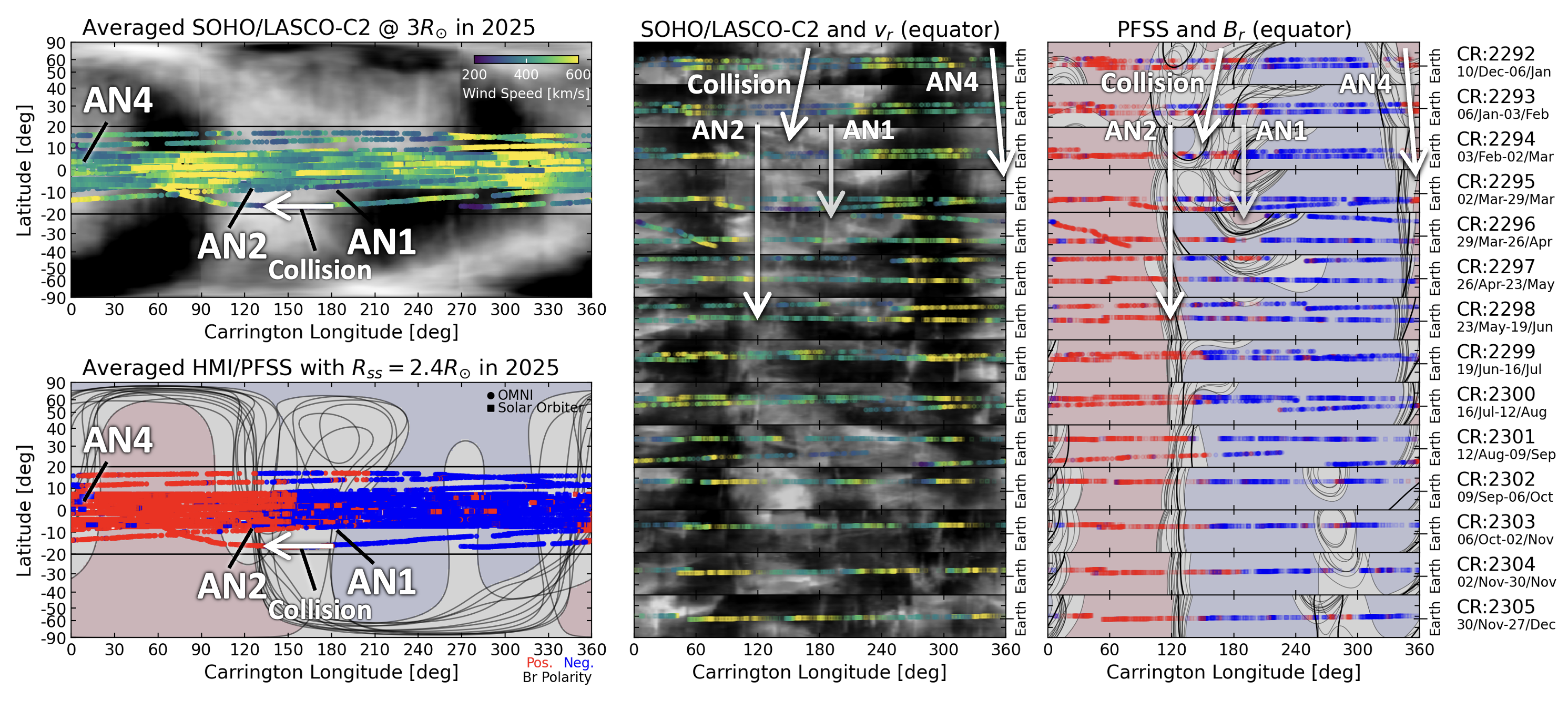}
    \caption{Solar wind and coronal magnetic field summary in Carrington coordinates for 2025. \textit{Top left}: Averaged SOHO/LASCO-C2 scattered white light distribution at $3R_{\odot}$, overlaid with hourly in situ solar wind speed from OMNI (at 1au) and Solar Orbiter, ballistically back-mapped to the source surface. \textit{Bottom left}: Coronal magnetic field polarity in the same style as Fig. \ref{fig:2024vs2025}, overlaid with hourly in situ radial magnetic field polarity from OMNI and Solar Orbiter. \textit{First column}: Time-evolution of the equatorial LASCO scattered white light and solar wind speed. \textit{Second column}: Time-evolution of equatorial coronal magnetic field polarity and solar wind magnetic field polarity. Within each CR, the coronal magnetic field polarity is compared across 14 ADAPT-GONG magnetograms spread equally throughout the CR. Time progresses from top to bottom in both columns, and the positions of AN1, AN2, and AN4 are marked with arrows. }
    \label{fig:anchorHCS}
\end{figure*}

AN1 and AN2 were separated by $\sim180^{\circ}$ longitude from the third active nest in both 2024 (AN3) and 2025 (AN4). Figure \ref{fig:2024vs2025} compares the spatial distribution of these active nests with the coronal magnetic field polarity. The orientation of the Sun's dipole magnetic field component remained stable across both years. In 2024, all three nests were located in the southern hemisphere. While each nest influenced the position of the HCS, AN1 and AN2 were the most effective at anchoring and inclining it. Following Hale's polarity law for solar cycle 25 \citep{hale1919magnetic}, the active regions within all three nests typically shared a consistent east-west magnetic field polarity. However, only AN1 and AN2 aligned with the pre-existing non-axisymmetric dipole field, whereas AN3 did not. Previous studies described the HCS above matching orientation nests as a Hale sector boundary and linked the underlying active regions to enhanced activity \citep[e.g.][]{loumou2018association}. By 2025, AN3 had dissipated and AN4 formed in the northern hemisphere. After re-emerging from the collision, AN2 was more active than AN1. The east-west magnetic field polarity of AN4, with a longitudinal separation of $\sim$180$^{\circ}$ from AN1 and AN2, now also aligned with the non-axisymmetric dipole field. Consequently, both AN2 and AN4 acted as Hale sector boundaries in 2025, with the nesting pattern reinforcing the non-axisymmetric dipolar component. This scenario parallels the rejuvenation of the Sun's large-scale magnetic field in 2014, where a southern active nest abruptly strengthened the non-axisymmetric dipole component \citep[see][]{sheeley2015recent}.

To explore how active nests influenced the heliosphere, we focused on the 2025 nesting pattern as this was the most clearly established in the solar corona. Figure \ref{fig:anchorHCS} combines a series of PFSS models from 2025, with the corresponding Carrington maps of scattered white-light observations from the Large Angle and Spectrometric Coronagraph \citep[LASCO;][]{brueckner1995large} on board the Solar and Heliospheric Observatory \citep[SOHO;][]{domingo1995soho}. These remote-sensing observations were compared to the hourly averaged in situ solar wind speed and radial magnetic field polarity taken from the OMNI database \citep{king2005solar} and Solar Orbiter using the Solar Wind Analyser \citep[SWA;][]{owen2020solar} and Magnetometer \citep[MAG;][]{horbury2020solar} instruments. For each CR, we computed 14 additional PFSS models using equally spaced ADAPT-GONG magnetograms to examine the variability within each solar rotation. The location of the HCS was tightly constrained in almost all CRs to lie above the active nests, with AN1 and AN2 separating after the collision and drifting to 180$^{\circ}$ and 120$^{\circ}$, respectively (see Appendix \ref{ap:2025_flares}). The HCS was located within $\sim20^{\circ}$ longitude of AN2 for 10 months in 2025.

The Sun's large-scale magnetic field was remarkably stable and dipolar in 2025, despite solar maxima being associated with higher-order magnetic field components \citep{derosa2012solar,finley2023evolution}.  This organisation resulted from the persistence of AN1 and AN2 over almost two years, whose emerging flux systematically strengthened the pre-existing non-axisymmetric dipole component of the Sun's magnetic field, while the polar magnetic field were near their weakest. The reservoir of positive-polarity flux east of AN2 (seen in epoch (d) of Fig. \ref{fig:collision}), together with the reconfigured large-scale field, facilitated the growth of a large coronal hole near 100° longitude, while the negative-polarity flux from AN1 fed into coronal holes further west. Figure \ref{fig:anchorHCS} shows low electron density (dark) regions in the LASCO/C2 coronagraph data associated with fast wind, with speeds of 600 to 800 km/s, separated by higher density (white) slow wind above the inclined HCS. The imbalanced magnetic fluxes surrounding the active nests also created filament channels (highlighted in Fig. \ref{fig:nestCollisionView}). As previously noted, the quadrupole-to-dipole energy ratio dropped from $\sim40$ in 2023 to near unity in 2025. A similar shift occurred in cycle 24, where the ratio decreased from $\sim100$ in 2012 towards unity in 2014 following the emergence of an active nest in the southern hemisphere \citep[see Fig. 3 in][]{finley2024nested}. These events demonstrate that active nests can act as topological anchors for the heliosphere, dictating the geometry of the solar wind long after the main flaring activity had subsided.

\subsection{Forecasting solar wind connection science}

\begin{figure}[h!]
    \centering
    \includegraphics[trim=0cm 0cm 0cm 0cm, clip, width=0.5\textwidth]{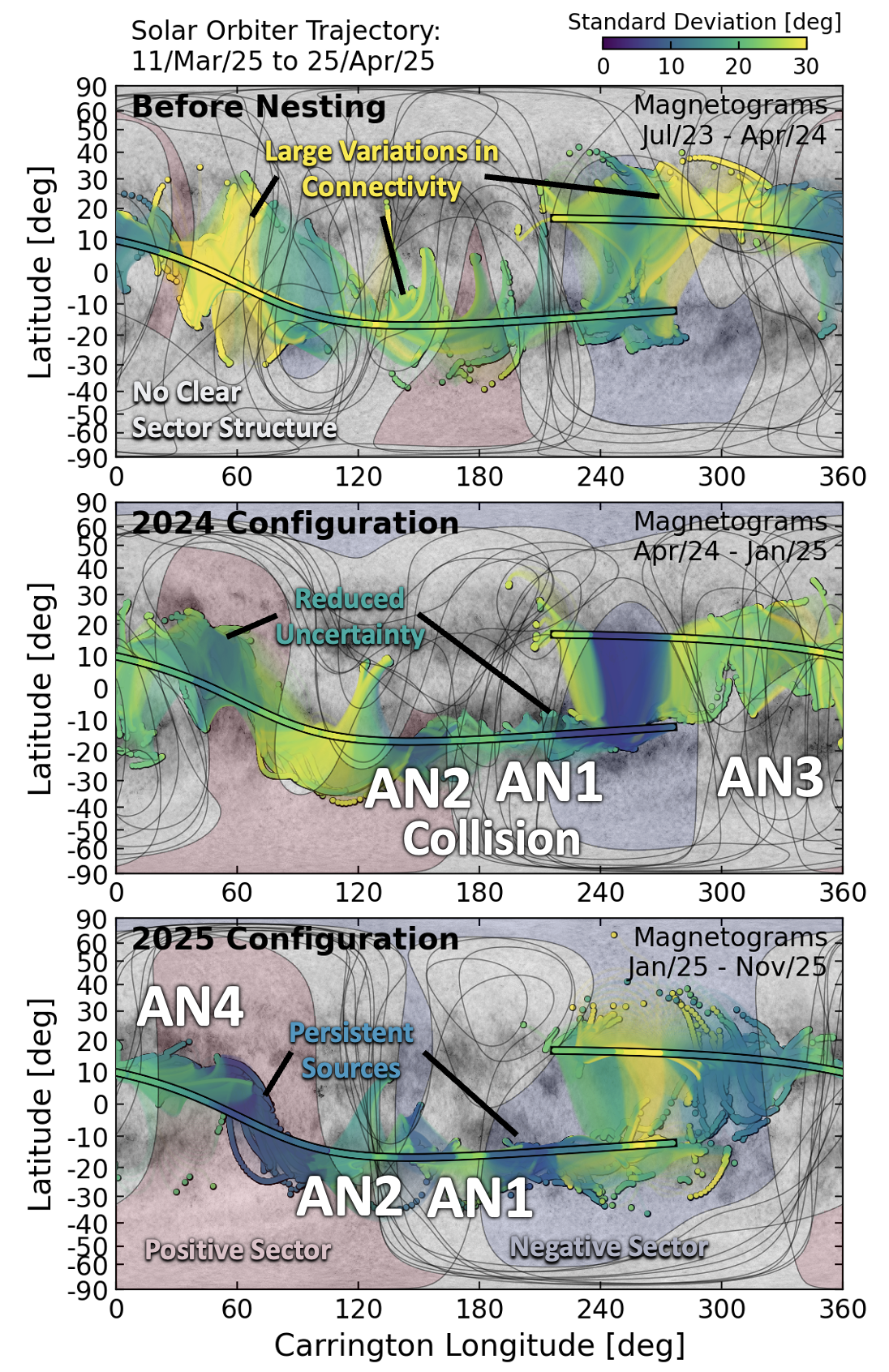}
    \caption{Solar wind connectivity across different coronal configurations, mapped along Solar Orbiter’s trajectory from 11 March to 25 April 2025. Connectivity is evaluated for three distinct epochs using ten consecutive SDO/HMI Carrington magnetograms each. \textit{Top}: Pre-nesting baseline period. \textit{Middle}: 2024 CRs featuring AN1, AN2, and AN3. \textit{Bottom}: 2025 CRs featuring AN1, AN2, and AN4. Panels show averaged SDO/AIA EUV emission, persistent heliospheric polarity sectors, and source region distributions. Trajectories, field lines, and source locations are coloured by their standard deviation across the ten magnetograms, where lower values indicate more stable connectivity with favourable sources.}
    \label{fig:connectivity}
\end{figure}

Solar Orbiter’s long-term planning is executed approximately six months prior to each observing window; this is required to align time-dependent data acquisition with available on-board storage and downlink capabilities. As the solar corona and wind were so heavily constrained by active nests in 2024 and 2025, we assessed the potential benefits this could have offered to mission level planning. We primarily focused on solar wind connection science \citep[e.g.][]{yardley2023slow}, as this requires linking remote-sensing observations to in situ solar wind measurements. The dominance of active nests at solar maximum ensured that the organisation of the fast and slow solar wind streams remained relatively self-similar in Carrington coordinates. This improved the reliability of connectivity tools that rely on PFSS and ballistic back-mapping techniques \citep[e.g.][]{rouillard2020models} by reducing the uncertainty introduced by the stochastic emergence of new active regions. 

We investigated how magnetic connectivity varied during a simulated Solar Orbiter observing campaign by systematically varying the nesting configuration. Three time periods were chosen: (1) July 2023 to April 2024, (2) April 2024 to January 2025, and (3) January 2025 to November 2025. These correspond to a baseline period before nesting in 2024, the nest configuration from 2024 (with AN1, AN2, and AN3), and the configuration from 2025 (when AN4 replaced AN3). For each epoch, we produced PFSS models from 10 consecutive Carrington magnetograms. We used the trajectory of Solar Orbiter from 11 March to 25 April 2025; the first inclined orbit after the Venus gravity assist in February 2025 (see Fig. \ref{fig:orbit2025}). During this time, the spacecraft passed directly over  the longitudes of AN1 and AN2. We focused on the varying coronal magnetic field topology and so assumed a constant 400 km/s back-mapping to the PFSS source surface when evaluating the connectivity.

Figure \ref{fig:connectivity} details how the solar wind connectivity varied during the three simulated observing campaign with Solar Orbiter. For each configuration, we evaluated the solar wind sources from all ten magnetograms and calculated the standard deviation of their mapped positions on the solar surface. Large variations (more than 20-30$^{\circ}$) in source locations signalled unpredictability, i.e. magnetograms from different CRs would not produce a similar forecast. The overall standard deviation in solar wind source locations decreased between each epoch as the coronal magnetic field became constrained by active nests and the tilted dipole topology; although significant dispersion remained around crossings of the HCS and pseudo streamers. The growth of the non-axisymmetric dipole established a clear magnetic field sector structure, as explored in Fig. \ref{fig:anchorHCS}. For the positive sector next to AN2 in 2025, the connectivity was extremely well determined; consequently, this area was subject to multiple connection science campaigns with Solar Orbiter during 2025 and 2026. The unipolar magnetic fields in these regions contributed to the formation and maintenance of coronal holes \citep[see][for similarities with the 2014 active nest]{yoshida2026temporal}. The stability of the magnetic connectivity around AN1 and AN2 was linked to their flux emergence and decay rates, with AN1 more active in 2024 and AN2 in 2025.

The presence of active nests created more favourable conditions for solar wind connectivity forecasting. In late 2025, magnetograms obtained several months in advance provided sufficient constraints for operational planning. However, we lack the ability to predict the initial emergence of active nests or the factors that govern their lifetimes. This is further complicated by unseen far-side flux emergence, which can influence the large-scale structure of the coronal magnetic field and solar wind \citep{perri2024impact, heinemann2025quantifying}. Probing the solar interior with helioseismology could be used to monitor active nests on the Sun's far-side \citep[][]{yang2024combined} or detect their feedback on large-scale flows \citep{sen2026active} and global oscillations \citep[][]{mehta2022cycle}. In the 2030s, ESA's Vigil mission \citep{west2025esa} will also provide valuable data to help monitor global solar activity. Numerical simulations remain indispensable for probing the mechanisms underlying dynamo magnetic field generation \citep[e.g.][]{brun2022powering}. Future studies could estimate the toroidal magnetic field reservoir accumulated at the base of the convection zone during the previous cycle \citep[][]{cameron2015crucial,finley2024well} to constrain the nesting lifetimes and track the toroidal flux budget as a function of longitude.

\section{Conclusions}
Combining multi-viewpoint observations from near-Earth satellites and Solar Orbiter, we investigated how nested flux emergence affected global magnetic activity during solar cycle maximum. We identified three active nests in the southern hemisphere that governed the vast majority of solar activity in 2024. Two active nests (AN1 and AN2) converged over several months, colliding in October around 180$^{\circ}$ Carrington longitude, whilst a third active nest (AN3) formed near 355$^{\circ}$. This region produced NOAA AR13664 in May, which caused the largest geomagnetic storm of the current solar cycle. The three active nests accounted for nearly 80\% of all flares over the entire Sun during the 156 days of nearly continuous monitoring (from April to October 2024) and roughly 70\% of all flares observed in 2024. This activity was maintained by recurring flux emergence within each nest. Complex flaring NOAA active regions (with $\beta\gamma$, $\beta\delta$, and $\beta\gamma\delta$ classifications) were more prevalent within the active nests than elsewhere on the Sun due to the interaction of pre-existing and emerging magnetic flux; nests accounted for 59\% of the 70 complex regions in 2024.

The convergence of AN1 and AN2 was driven by the migration of flux emergence sites towards one another, with successive emergences shifting by $4-10^{\circ}$ in longitude on timescales of 12 to 18 days. Additionally, the rapid flux emergence that formed AR13664 in May appeared to locally suppress subsequent activity, restricting the following flux emergence events in AN3 to the perimeter of the initial region. Taken together, these behaviours suggest that nested magnetic activity is rooted in coherent sub-surface magnetic field reservoirs. This interpretation is further supported by the spatial distribution of flux emergence in 2025. Two nests re-emerged from the remnants of the AN1 and AN2 collision, shifted by $\sim30^{\circ}$ in longitude relative to the original nests (see Fig. \ref{fig:CRsummary2025south}). Meanwhile, AN3 disappeared, and a new active nest (AN4) formed at the same longitude but in the northern hemisphere (see Fig. \ref{fig:CRsummary2025north}). The overall distribution of flux emergence in 2025 largely mirrored the quiet areas from 2024. 

The two main nesting sites in 2025 (AN2 and AN4) had a longitudinal separation of 180$^{\circ}$ and east-west magnetic field polarities that reinforced the tilted dipole configuration in the coronal magnetic field and solar wind. The resulting heliospheric magnetic field sector structure was stable for 10 months, with the HCS consistently located above AN2 and AN4 (each within $\sim 20^{\circ}$). Unipolar magnetic flux diffusing from the leading and trailing edges of the active nests contributed to the formation of large coronal holes that launched fast (600-800 km/s) solar wind streams into the heliosphere. The persistence of this coronal topology over several solar rotations created more favourable target regions for solar wind connectivity studies.

The concentration of activity into discrete nests suggests that the underlying solar dynamo has a preference for flux emergence around certain locations in Carrington coordinates, even at the busiest time of the solar cycle. While the origins of the extended sub-surface magnetic flux reservoirs required to sustain nesting requires further investigation, the episodic migration of flux emergence events hints that near-surface convection could be modulating the emergence of magnetic flux organised within the deep interior. These findings offer a promising pathway towards improving long-term space weather forecasting. Crucially, active nests could be integrated into predictive models as recurring, quasi-predictable drivers of global space weather. However, bridging the gap between identifying active nests and forecasting either their initial emergence or their energetic flare production remains an objective for future research.

\begin{acknowledgements}
We thank Manuela Temmer for insightful feedback that helped contextualise this study with respect to previous solar cycles.
AJF, ASHT, and HE acknowledge support through the European Space Agency (ESA) Research Fellowship in Space Science.
This research has received funding from the European Research Council (ERC) under the European Union’s Horizon 2020 research and innovation programme (grant agreements No 810218 WHOLESUN and No 101125367 ExoMagnets), in addition to funding by the Centre National d'Etudes Spatiales (CNES) Solar Orbiter project, the French Agence Nationale de la Recherche (ANR) project STORMGENESIS \#ANR-22-CE31-0013-01, and the Institut National des Sciences de l'Univers (INSU) via the Action Thématique Soleil-Terre (ATST).
Solar Orbiter is a mission of international cooperation between ESA and NASA, operated by ESA.
Data supplied courtesy of the SDO/HMI and SDO/AIA consortia. SDO is the first mission to be launched for NASA's Living With a Star (LWS) Program.
Data manipulation was performed using the numpy \citep{2020NumPy-Array}, scipy \citep{2020SciPy-NMeth}, and pySHTOOLS \citep{wieczorek2018shtools} python packages.
Figures in this work are produced using the python package matplotlib \citep{hunter2007matplotlib}.
\end{acknowledgements}

%
%

\bibliographystyle{aa}
\bibliography{adam}

\onecolumn
\begin{appendix}

\section{Assigning GOES classifications to solar flares observed by STIX}
\label{ap:goes_stix}

\begin{figure}[htbp]
    \centering
    \includegraphics[trim=0cm 0cm 0cm 0cm, clip, width=0.6\textwidth]{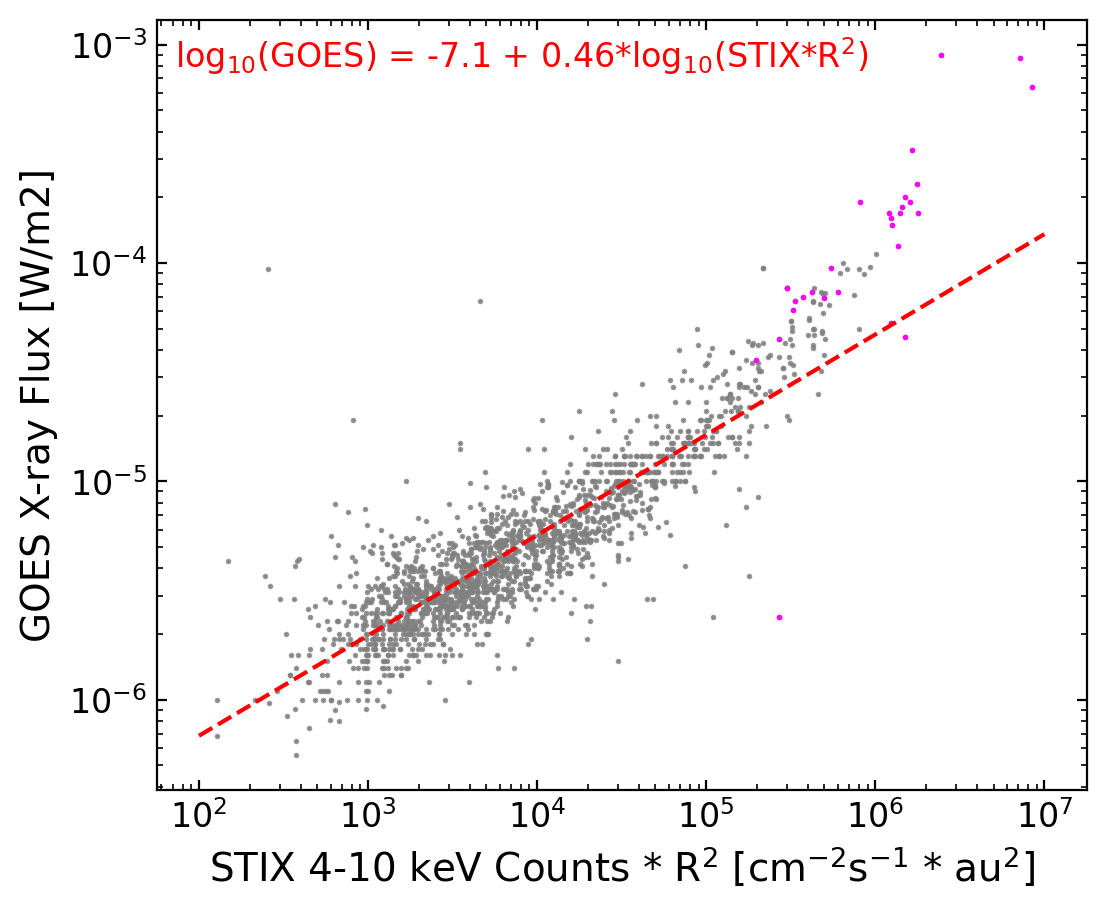}
    \caption{Comparison of GOES 1.5--12.4 keV peak flux versus Solar Orbiter STIX 4--10 keV counts for flares observed simultaneously at the beginning and end of 2024. Flares that triggered the insertion of the STIX attenuator are coloured magenta. The fit of Eq. (\ref{eq:flare}) to the flare distribution (parameters shown in the upper left) is plotted as a dashed red line.}
    \label{fig:goes_stix}
\end{figure}

To compare flares observed with Solar Orbiter STIX on the same scale as those measured near-Earth by GOES, we constructed an empirical relationship between STIX counts and GOES flux, in the form of Eq. (\ref{eq:flare}). This relation has two fit parameters, $a$ and $b$. As done previously in \citet{finley2025prolific} for 2022, we calibrated this relationship using flares observed by both instruments at the beginning and end of 2024. During these times, Solar Orbiter was closer to Earth with reduced viewing angle effects. Figure \ref{fig:goes_stix} displays the relationship between the two measurements. Flares that triggered the STIX attenuator to be inserted were removed from the fitting process. The dashed line shows the fit with parameters of $a=-7.1$ and $b=0.46$. 

\newpage
\section{Active nest collision as seen in EUV wavelengths}\label{ap:nestCollisionView}

\begin{figure*}[htbp]
    \centering
    \includegraphics[trim=0cm 0cm 0cm 0cm, clip, width=0.9\textwidth]{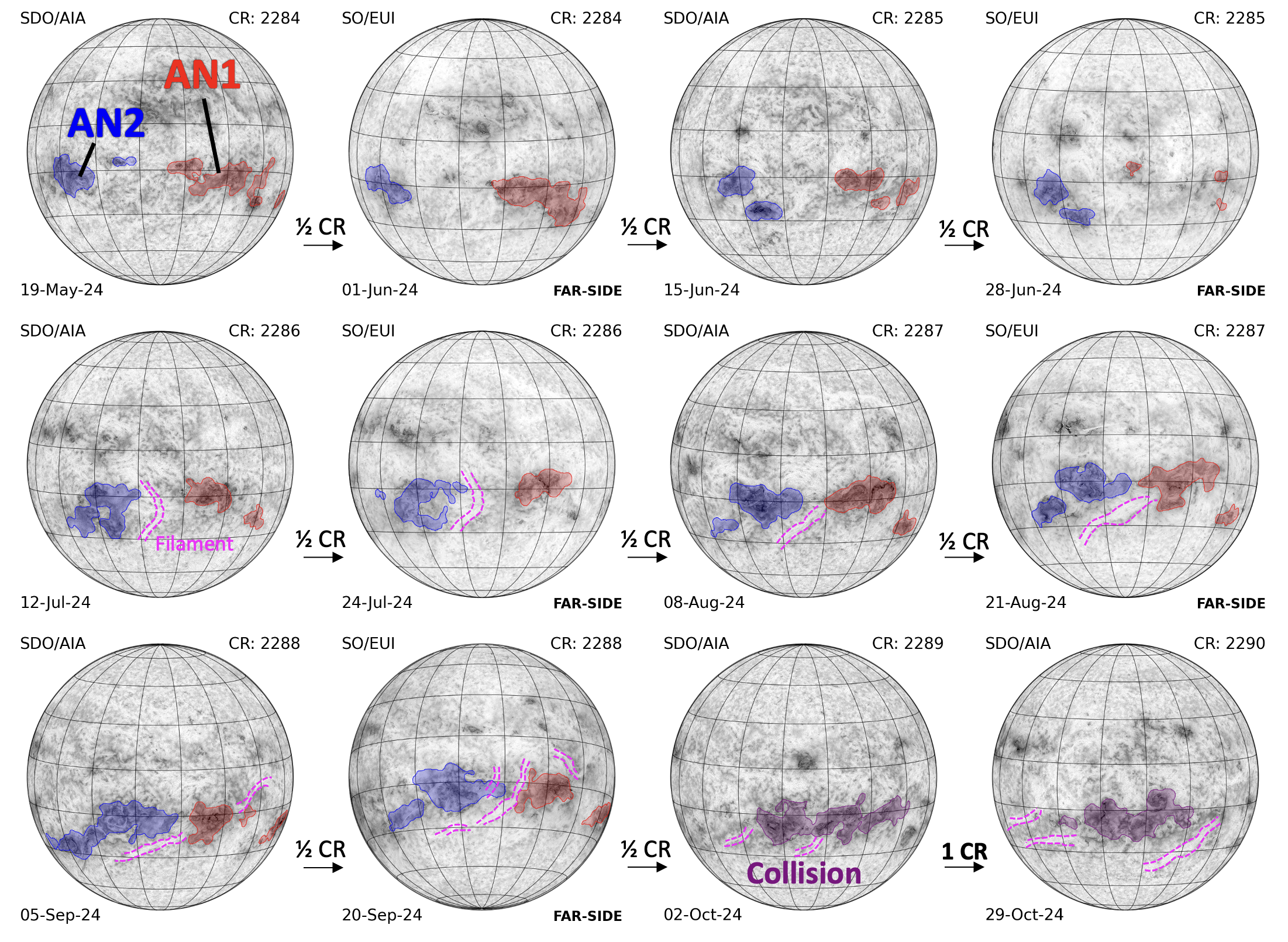}
    \caption{Nearly continuous monitoring in EUV 304~\AA\  of active nests AN1 and AN2 from May to October 2024 (grey scale). Smoothed intensity contours are overplotted to illustrate the convergence and collision of the two nests. The sequence of full-disk images alternates between AIA and EUI (when observing the Sun's far side) with approximately two weeks separating each panel. Filament channels are highlighted with dashed magenta lines.}
    \label{fig:nestCollisionView}
\end{figure*}

Figure \ref{fig:nestCollisionView} shows a sequence of full-disk EUV images spanning the nearly continuous monitoring window of 2024. In each image, the location of AN1, AN2, or their collision product are highlighted, along with filament channels that formed around the active nests. The latitude of Solar Orbiter varied by around 10$^{\circ}$ over this period, creating different viewing angles compared to Earth. The final two images in the sequence are from AIA and show the combined contour once AN1 and AN2 converged. This occurred as Solar Orbiter returned towards Earth and so we only present the observations from AIA.

\newpage
\section{Identifying active nests in 2025} \label{ap:2025_flares}

\begin{figure}[htbp]
    \centering
    \includegraphics[trim=0cm 0cm 0cm 0cm, clip, width=\textwidth]{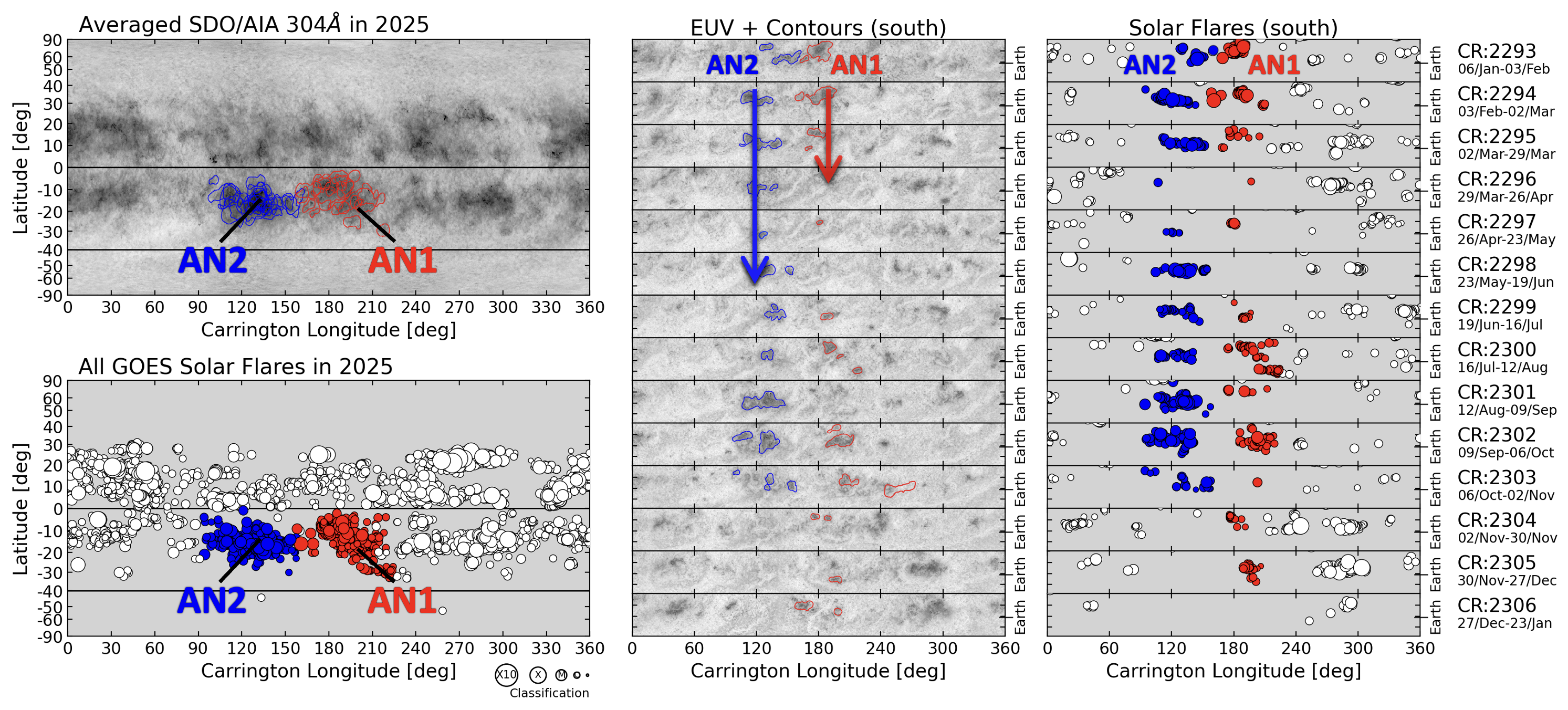}
    \caption{Same as Fig. \ref{fig:CRsummary} but now for the southern hemisphere in 2025. The re-emerging active nests AN1 and AN2 are highlighted in red and blue, respectively.}
    \label{fig:CRsummary2025south}
\end{figure}

\begin{figure}[htbp]
    \centering
    \includegraphics[trim=0cm 0cm 0cm 0cm, clip, width=\textwidth]{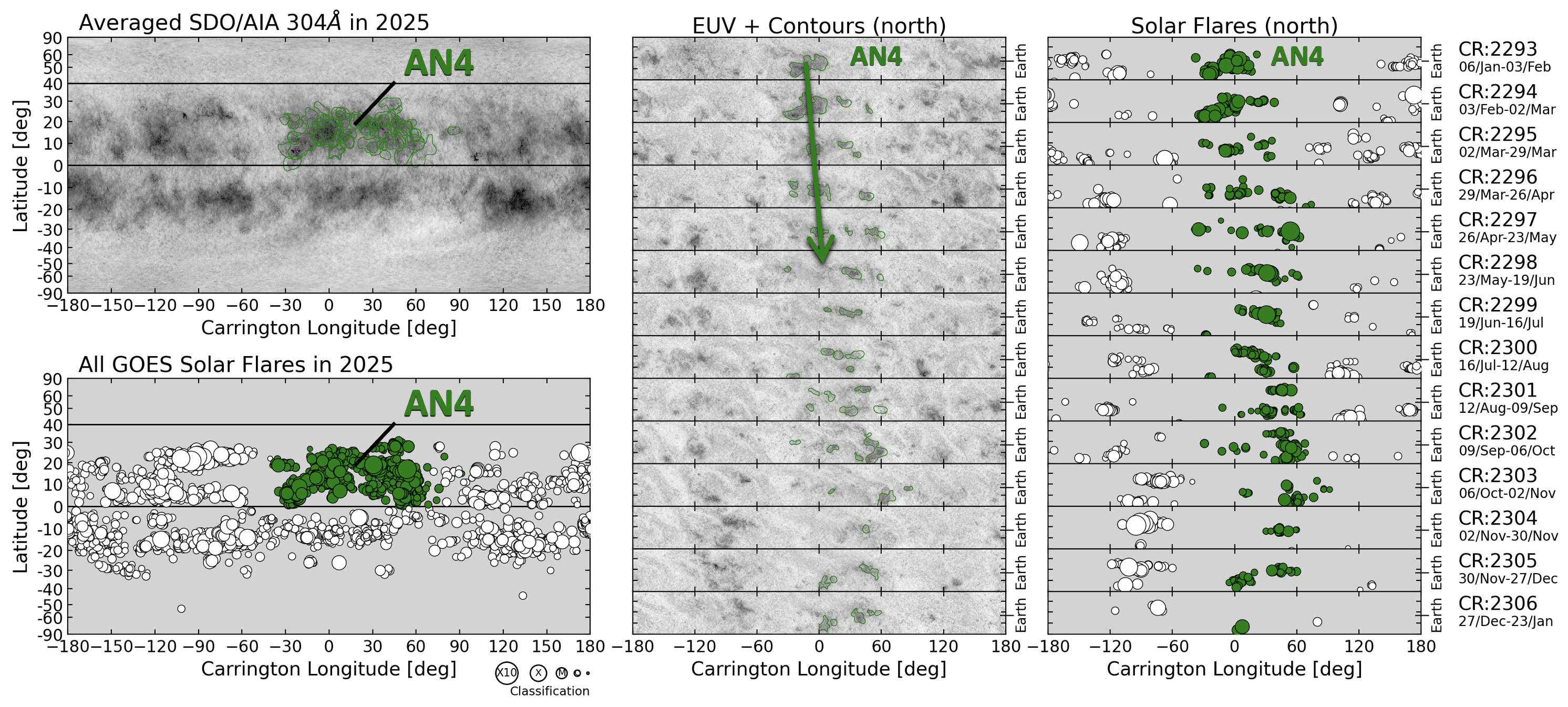}
    \caption{Same as Fig. \ref{fig:CRsummary2025south} but now for the northern hemisphere. The longitude axis has been shifted by 180$^{\circ}$ to better highlight AN4.}
    \label{fig:CRsummary2025north}
\end{figure}

We focused on the Sun's flaring activity at solar maximum in 2024; however, nesting continued into 2025. Figures \ref{fig:CRsummary2025south} and \ref{fig:CRsummary2025north} highlight three active nests from 2025, two in the southern hemisphere and one in the northern hemisphere. Each nest was tracked following the methodology outlined in Sect. \ref{sect:nests}. After the collision of AN1 and AN2 in October 2024, the migrating pattern of flux emergence separated back into two distinct regions again in January 2025. The collision product persisted from CR2289 to CR2292 (see Fig. \ref{fig:CRsummary}). In Fig. \ref{fig:CRsummary2025south} we identify the two nests and conserve the labelling convection from 2024 of AN1 and AN2, given the clear similarities. However, these regions were notably weaker and shifted by $\sim 30^{\circ}$ Carrington longitude with respect to their 2024 counterparts. In the northern hemisphere, Fig. \ref{fig:CRsummary2025north} shows a new active nest at 10-20$^{\circ}$ Carrington longitude. 

\begin{figure}[htbp]
    \centering
    \includegraphics[trim=0cm 0cm 0cm 0cm, clip, width=0.5\textwidth]{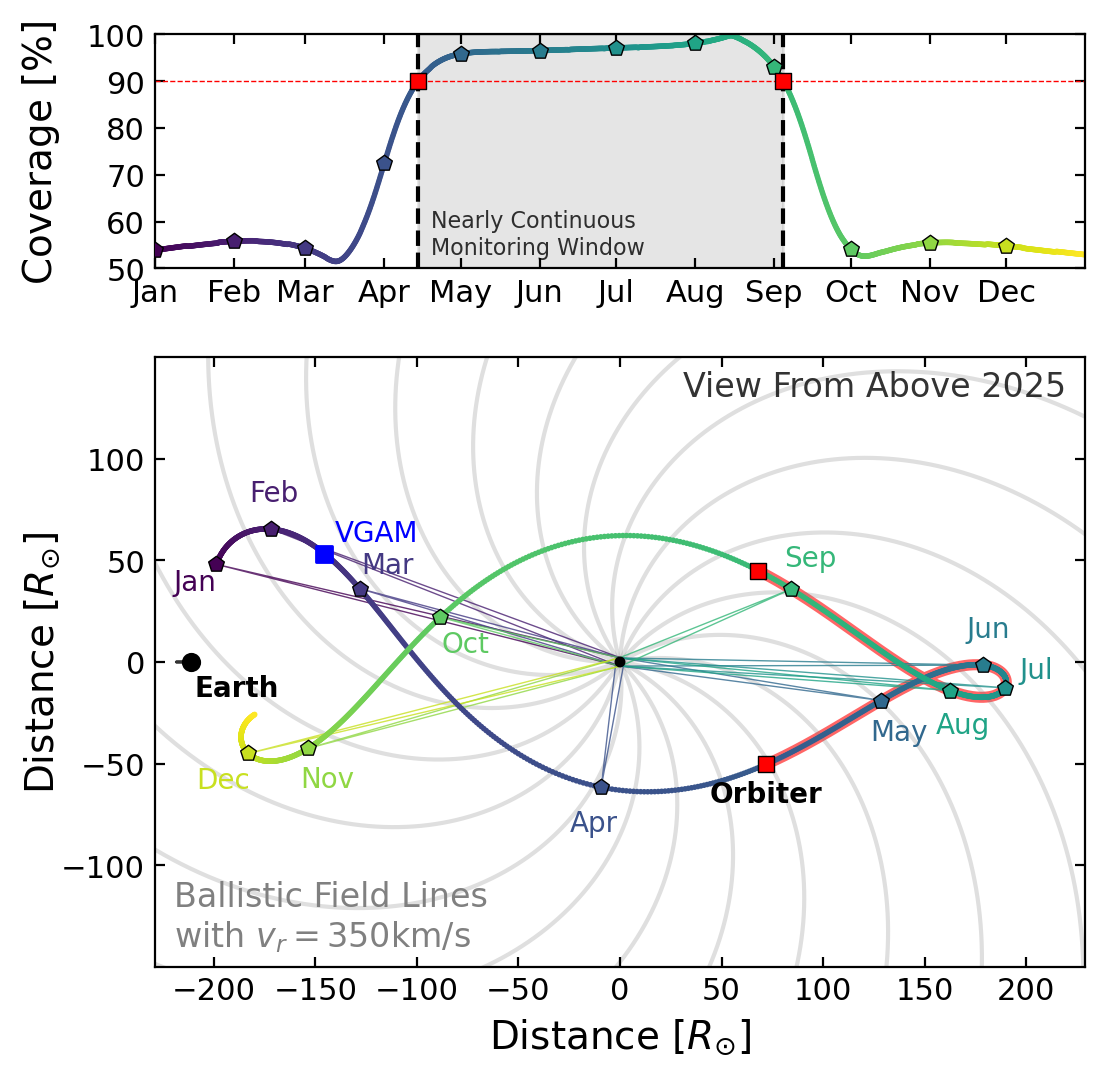}
    \caption{Same as Fig. \ref{fig:soloOrbit} but now for 2025. The 90\% coverage window spanned 143 days. The Venus gravity assist manoeuvrer (VGAM) in February 2025 is marked by a blue square.}
    \label{fig:orbit2025}
\end{figure}

The nesting of flux emergence in 2025 is a compelling target for further research. Figure \ref{fig:orbit2025} highlights the multi-viewpoint coverage available for such an analysis using Solar Orbiter and near-Earth satellites. The nearly continuous monitoring window in 2025 lasted for 143 days. From CR 2303 onwards, the longitudinal distribution of flux emergence diverged from the nesting pattern characterised in this work, triggering a shift in the coronal magnetic field topology. The dipole axis, previously fully inclined to the rotation axis by the active nests, rapidly advanced towards the poles. At the same time, the Sun's large-scale magnetic field became more quadrupolar in appearance.

\end{appendix}
\end{document}